\documentclass{emulateapj}
\usepackage{graphics}
\newcommand{\en}{{\cal N}}
\newcommand{\period}{{P}}
\newcommand{\flux}{{\cal F}}
\newcommand{\losscone}{{\rm lc}}
\newcommand{\beq}{\begin{equation}}
\newcommand{\eeq}{\end{equation}}
\newcommand{\bea}{\begin{eqnarray}}
\newcommand{\eea}{\end{eqnarray}} 
\newcommand{\accel}{{\cal A}}
\newcommand{\quinlanj}{{\cal J}}
\newcommand{\rootr}{{j}}
\newcommand{\rootrlc}{{\rootr_\losscone}}
\newenvironment{inlinefigure}{
\def\@captype{figure}
\noindent\begin{minipage}{0.95\linewidth}\begin{center}}
{\end{center}\end{minipage}\smallskip}

\begin{document}
\title{Long Term Evolution of Massive Black Hole Binaries}
\author{Milo\v s Milosavljevi\' c $^{1,2}$ and David Merritt$^1$}
\affil{$^1$ Department of Physics and Astronomy, Rutgers University,
New Brunswick, NJ 08903; \\
$^2$ Theoretical Astrophysics, California Institute of Technology,
Pasadena, CA 91125; \\
milos@tapir.caltech.edu, merritt@physics.rutgers.edu
}
\righthead{LONG-TERM EVOLUTION OF MASSIVE BLACK HOLE BINARIES}
\lefthead{MILOSAVLJEVI\'C \& MERRITT}
\begin{abstract}
The long-term evolution of massive black hole binaries at the centers of 
galaxies is studied in a variety of physical regimes,
with the aim of resolving the ``final parsec problem,''
i.e. how black hole binaries manage to shrink to separations
at which emission of gravity waves becomes efficient.
A binary ejects stars by the gravitational slingshot
and carves out a loss cone in the host galaxy.
Continued decay of the binary requires a refilling of the loss cone.
We show that the standard treatment of loss cone refilling,
derived for collisionally relaxed systems like globular clusters, can
substantially underestimate the refilling rates in galactic nuclei.
We derive expressions for non-equilibrium
loss-cone dynamics and calculate time scales for the decay of
massive black hole binaries following galaxy mergers, obtaining
significantly higher decay rates than heretofore.
Even in the absence of two-body relaxation, decay of binaries can
persist due to repeated ejection of stars returning to
the nucleus on eccentric orbits.
We show that this recycling of stars leads to a gradual, 
approximately logarithmic dependence of the binary binding energy on time.
We derive an expression for the loss cone refilling induced by 
the Brownian motion of a black hole binary.
We also show that numerical $N$-body
experiments are not well suited to probe these mechanisms over long times
due to spurious relaxation. 
\end{abstract}

\keywords{black hole physics --- galaxies: nuclei -- stellar dynamics}

\section{Introduction}
Larger galaxies 
grow through the agglomeration of smaller galaxies and
protogalactic fragments.  
If more than one of the fragments contains a massive black hole (MBH), 
the MBHs will form a bound system in the merger product
\citep{Begelman:80,Roos:81,Valtaoja:89}.
This scenario has received considerable attention because the 
ultimate coalescence of such a pair would
generate an observable outburst of gravity waves \citep{thb76}.
\citet{Begelman:80} showed that the evolution of a MBH
binary can be divided into three distinct phases: 
1. As the galaxies
merge, the MBHs sink toward the center of the new galaxy 
via dynamical friction \citep{cha43} where they form a binary.  
2. The binary
 continues to decay via gravitational slingshot interactions
\citep{sva74} in which
 stars on orbits intersecting the binary's are ejected at velocities
 comparable to the binary's orbital velocity, while the binary's binding energy
 increases. 
3. If the binary's
 separation decreases to the point where the emission of gravity waves becomes
 efficient at carrying away the last remaining angular momentum, 
the MBHs coalesce rapidly.  In this paper we explore Phase 2
 and its transition to Phase 3; 
this transition is understood to be the bottleneck of
 a MBH binary's path to coalescence.

 The picture of \citet{Begelman:80} has remained essentially valid in the face of several important developments in intervening years.  Early 
theoretical studies
 envisioned the centers of galaxies as constant-density cores, with small,
 embedded stellar cusps that develop around central MBHs via
 collisional relaxation \citep{pee72,baw76,Young:77,Ipser:78}.  
 Observations at the time lacked the
 resolution to verify this picture.
In the past decade, space-based observations have revealed
 that, in the majority of early-type galaxies,
the central luminosity density increases continually inward as an
 approximate power law $\rho(r)\sim r^{-\gamma}$ down to regions where the MBH dominates the gravitational force \citep{fer94,geb96,res01}.  
The typical luminosity profile of a faint elliptical galaxy is a nearly-featureless power law with
 $1.5\le\gamma\le2.5$ at the resolution limit.
The times required to build cusps
 via two-body relaxation are longer than the age of
 the Universe in most nuclei (e.g.~\citealt{fab97}) and so relaxation
 is not a viable explanation for how the cusps formed.
 In this work we assume that galaxies inherited their steep central density profiles from an early epoch in which the MBH grew.  Cold dark matter (CDM) halos produced by hierarchical structure formation that presently host galaxies and MBHs were cuspy at the time of virialization (e.g.~\citealt{Ghigna:00,Power:03}). 
On smaller scales, the stellar cusps may have resulted from gaseous dissipation of star-forming medium inside pre-existing CDM cusps, from adiabatic response of the stars to growth of the MBH, or both.
  The same processes that led to cusp formation
 were likely to have fueled the early growth of MBHs and, thus, set the
 MBHs and galaxies on a course of co-evolution.  We therefore assume
 that the loci of MBHs have coincided 
with the centers of cosmic overdensities (dark matter and stellar
 cusps) since the time MBHs first ascended into the massive range,
 $M_\bullet\sim10^6M_\odot$.  

The cusps that accompany the MBHs endow
 them with an enlarged effective dynamical mass.  
\citet{Milosavljevic:01} found that
the dynamical friction time scale
 (Phase 1) is a function of this effective mass and much shorter than the orbital decay time implied by the masses of the MBHs alone.
This is of critical importance for the capacity of MBHs originating in
 different, merging galaxies to capture each other and 
form a hard binary in the resulting galaxy.  
A MBH binary is customarily called
``hard'' if its binding energy per
 unit mass exceeds the specific kinetic 
energy of ambient stars, $G\mu/4a\gtrsim\sigma^2$,
 where $\mu=M_1M_2/(M_1+M_2)$ is the reduced mass,
$a$ is the binary's semi-major axis and
 $\sigma$ is the 1D stellar velocity dispersion.
\citet{Milosavljevic:01} also found that in mergers of
 equal-mass $\rho(r)\sim r^{-2}$ galaxies, 
the true separations at which the binaries settle at the beginning of Phase 2 
are $\sim 10$ times smaller
than implied by arguments based on the equipartition of energy.  

Masses of MBH's correlate
exceptionally well with the central line-of-sight velocity dispersions of their host galaxies \citep{fem00,geb00}.  
It is likely that the 
masses are also related to the depth of 
the potential wells associated with
the dark matter halos in which the 
host galaxies are embedded \citep{Ferrarese:02a}, 
which justifies attempts to characterize the 
{\it evolution} of the MBH population using the statistics of
hierarchical formation of structure in dark matter cosmologies \citep{kah00,Menou:01,Volonteri:02}.  As a
corollary, one can estimate the frequency of MBH binary 
coalescence at redshifts that are beyond the reach of existing
telescopes but within reach of a future 
space-based gravity wave detector such as LISA.\footnote{Laser Interferometer
  Space Antenna, http://lisa.jpl.nasa.gov} The coalescence of two
$10^5-10^6 M_\odot$ MBHs occurring practically anywhere in the observable Universe would generate a significant signal
for LISA.

Early calculations (e.g.~\citealt{Haehnelt:94}) 
identified the MBH coalescence event rate with the galaxy merger rate.  It was
subsequently noted that stellar-dynamical processes are inefficient at
facilitating the decay of MBH binaries when their separations shrink
below about one parsec and that the ultimate 
coalescence should not be taken for
granted \citep{val96,gor00,Milosavljevic:01,Yu:02a}.  
Circumstantial evidence, however,
suggests that long-lived MBH binaries are
in fact rare.  
The coincidence of mean black hole mass densities derived from kinematical
studies of local galaxies and distant quasars 
(\citet{Ferrarese:02b}) suggests that MBHs are rarely ejected from galaxy centers by the strong interactions that would accompany the infall of a third MBH into a nucleus containing an uncoalesced binary.
Furthermore there is evidence for MBH coalescences in the morphology of radio galaxies, at a rate approximately equal to the galaxy merger rate 
\citep{MerrittEkers:02}.

There are many processes that could  
drive the MBH binaries that form in galaxy mergers to coalescence.  
In this paper we present some new ideas regarding one
such process, the gravitational slingshot ejection of stars.

Ejection of stars from the galaxy nucleus
will result in a depletion of stars in the ``loss cone.''
We define the loss cone as the phase space domain in which orbits
approach the binary closely enough to be influenced by the near field of
the individual binary components.  
The orbits in the loss cone are subject to
gravitational capture and slingshot ejection.  The loss cone has thus
traditionally been viewed as a sink in the phase space, surrounded by a
distribution of stars in equilibrium with respect to two-body gravitational
encounters.  The latter assumption allows a steady-state solution for the
phase space density near the loss cone boundary to be derived (e.g.~\citealt{cok78}); this steady state reflects a balance between the supply of stars to the loss cone by gravitational scattering and the loss due to capture or ejection.
In this standard picture, stars that scatter into the loss cone are
ejected from the system with net energy loss, resulting in a more
tightly bound binary.  
After all of the stars originally inside the loss cone have
been removed, continued decay of the binary hinges on the flux of stars
scattering into the loss cone.  

In this paper we study the physics of the ``worst case scenario,'' involving
nearly spherical galaxies, in which the MBH binaries
are least likely to coalesce in their lifetime.   
In \S~\ref{sec:reviewck} we review the standard solution for the flux.  
In \S~\ref{sec:issues} we question the applicability of some of the
assumptions that were made to derive that solution.  
In \S~\ref{sec:timedep} and
\S~\ref{sec:secondary} we introduce some novel aspects of the dynamics of stars
in the loss cone and outline an iterative procedure for calculating the
long-term evolution of a massive MBH binary in a spherical
galaxy.  The details emphasized here tend to
shorten the timescale for the decay of MBH 
binaries, sometimes significantly. 
In \S~\ref{sec:nbody} we evaluate the effectiveness of $N$-body
simulations to model the long-term evolution of MBH binaries.  
In \S~\ref{sec:brownian} we address the implications of the Brownian
motion of MBH binaries for the long-term evolution.  In \S~
\ref{sec:coalescence} we review the likelihood for the coalescence of MBH
binaries.  
Extensions to axisymmetric and triaxial galaxies, 
as well as applications to specific galaxy models, 
are deferred for future work.

\section{Review of Standard Loss Cone Theory}
\label{sec:reviewck}
The theory of loss cone structure was originally
developed to model the rates of tidal disruptions of stars by 
MBHs \citep{Frank:76}.
Results reviewed here were derived in the
context of globular cluster-like stellar systems containing
MBHs \citep{lis77,cok78,Ipser:78} where the stars are tidally
disrupted after coming within the tidal radius of the MBH:
\beq
\label{eq:tidalradius}
r_{\rm tidal}=\left(\eta^2\frac{M_\bullet}{m_*}\right)^{1/3} r_* ,
\eeq       
where $m_*$ and $r_*$ are, respectively, the stellar mass and the
stellar radius, and $\eta$ is a form factor of order unity.
Recently, the loss cone paradigm has been utilized to
estimate rates of disruptions of stars by MBHs residing in
galaxy nuclei \citep{Sigurdsson:97,Syer:99,mat99}, 
as well as to calculate the rates of scattering of stars into
the capture zone of MBH binaries \citep{Yu:02a}.  
At the heart of these studies
lies the assumption that stellar systems, such as globular clusters or small
galaxies, are old compared with local two-body relaxation times.  This
justifies the description of these systems in terms of 
the time-independent Fokker-Planck equation \citep{Rosenbluth:57,cok78}.

For simplicity we restrict attention to
spherical galaxies; there the phase space distribution is a function of two
orbital integrals---the energy $E$ and the angular momentum $J$ of a star.
The Boltzmann equation reads: 
\beq
\label{eq:timeindfp}
\frac{\partial f}{\partial r} = \frac{1}{v_r} 
\frac{\partial}{\partial R} 
\left[\frac{\langle(\Delta R)^2\rangle}{2R} R
\frac{\partial f}{\partial R}\right] ,
\eeq
(e.g.~\citealt{cok78}, eq. 30), 
where $f(E,R,r)d^3rd^3v$ is the probability of finding a star in the phase
space volume element $d^3rd^3v$,
$v_r$ is the radial velocity of the star,
$R=J^2/J_c^2(E)$ is the square angular momentum in
units of the angular momentum of a circular orbit at energy $E$, 
and $\langle(\Delta R)^2\rangle$ is the diffusion coefficient
associated with the orbital integral $R$.  
At no risk of confusion we will 
variably refer to $R$ as ``angular momentum.''
Note that all terms representing diffusion in energy have been discarded; this
is justified because the gradients in $E$ are minuscule compared with the
gradients in $R$ in the vicinity of the loss-cone boundary.  The
diffusion coefficient scales inversely with the number of stars in the
galaxy:
\beq
\langle(\Delta R)^2\rangle\propto N^{-1} \sim \frac{m_*}{M_{\rm
    galaxy}} .
\eeq

Conditions for the validity of the time-independent,
orbit-averaged Fokker-Planck equation can be summarized as follows:
\beq
\period(E) \ll T_{\rm relax}(E) \ll T_{\rm age} ,
\eeq
where $\period(E)$ is a typical orbital period (or the crossing time) of a
star at energy $E$, $T_{\rm relax}$ is the local two-body relaxation time, and
$T_{\rm age}$ is the total lifetime of the system.  If the first inequality fails, perturbative
expansion of the Boltzmann equation is no longer valid, large deflections
dominate, and the system is in a gaseous phase.  
If the second inequality fails,
the system need not be collisionally relaxed and can keep evolving; then the
time-dependent Fokker-Planck equation is necessary.  We argue in the next
section that the second inequality is almost never satisfied in galaxies with
MBH binaries.

Let $r_{\rm bin}$ denote the distance from the binary's center of mass 
within which stellar orbits are strongly perturbed by 
the rotating quadrupole gravitational field of the binary.  
This distance is similar to the semi-major axis of the binary, 
$r_{\rm bin}\sim a$.
Stars that transgress a region around the binary $r\leq r_{\rm bin}$ are
subject to gravitational slingshot and are ejected from the system.  This
imposes a boundary condition on equation (\ref{eq:timeindfp}) of the form: 
\beq
\label{eq:bdrycond}
f(E,R,r)=0\ \ \ \ \ \ (r \leq r_{\rm bin}) .
\eeq
\citet{lis77} derived an orbit averaged solution $f(E,R)$ to equation
(\ref{eq:timeindfp}) subject to the boundary condition expressed by equation 
(\ref{eq:bdrycond}) to find:
\beq
f(E,R)\equiv \period^{-1}(E)\oint \frac{dr}{v_r} f(E,R,r)
= C \ln\left(\frac{R}{R_0}\right) ,
\label{eq:LS}
\eeq
where $C$ is a normalization factor
to be determined, and $R_0(E)$ is an energy-dependent
angular momentum cutoff.  

Equation (\ref{eq:LS}) assumes that $f(E,R,r)$ vanishes identically
when $R\leq R_0$.  
In the absence of two body relaxation, $R_0=R_\losscone$,
 where $R_\losscone(E)$ is the angular momentum of a particle with pericenter
 distance at the very edge of the sphere of capture,
$r_-(E,R_\losscone)=r_{\rm bin}$.  In the presence of relaxation, some
stars will enter 
{\it and} exit the consumption zone $R<R_\losscone$ all within
one orbital period, thereby evading consumption.  \citet{cok78} carried out
a boundary-layer analysis to find the following relation between $R_0$ and
$R_\losscone$:
\beq
R_0(E) = R_\losscone(E) \times 
\cases{ \exp(-q), & $q(E) > 1$ \cr
  \exp(-0.186 q -0.824 \sqrt{q}), & $q(E) < 1$} ,
\eeq
where $q(E)$ is the ratio of the orbital period at energy $E$ to 
the timescale for diffusional refilling of the consumption
zone at this energy:
\beq
q(E)\equiv \frac{1}{R_\losscone(E)} \oint 
\frac{dr}{v_r}
\lim_{R\rightarrow 0}
\frac{\langle(\Delta R)^2\rangle}{2R} . 
\eeq
When the period is much shorter than the refilling time scale ($q\ll 1$), 
the system is in a ``diffusive'' regime, the loss cone is largely empty and
$R_0\approx R_\losscone$.
When the period is much longer than the refilling time scale ($q\gg 1$), the
system is in a ``pinhole'' regime, the loss cone is largely full and $R_0\ll
R_\losscone$.  The energy at which $q=1$ is called the critical energy.
If the loss cone is as narrow as in the tidal disruption
problem, parts of the phase space at the largest energies 
$E\gtrsim 2\sigma^2$ will
be in the diffusive regime, while the less bound regions will be in the pinhole
regime.  Most of the lost cone flux originates near the critical energy.

The normalization factor $C$ can be expressed in terms of the isotropized
distribution function $\bar f(E)=\int_0^1 f(E,R) dR$ to obtain:
\beq
C(E)=\left(R_0-\ln R_0 -1 \right)^{-1} \bar f(E) ,
\eeq
implying that the orbit-averaged solution for the time-independent
Fokker-Planck equation reads (using $R_0\ll 1$):
\beq
\label{eq:soltimeind}
f(E,R)= \frac{\ln(R/R_0)}{\ln(1/R_0)-1} \bar f(E) .
\eeq
In this case, the number flux of stars into the loss cone is given by:
\bea
\label{eq:fluxtimeind}
\flux(E)dE &=& 4\pi^2 J_c^2  
\left\{
\oint \frac{dr}{v_r}
\lim_{R\rightarrow 0}
\frac{\langle(\Delta R)^2\rangle}{2R} 
\right\}
\nonumber\\ & &\times
\frac{\bar f}{\ln(1/R_0)-1}dE ,
\eea
where all quantities on the right hand side
depend on energy, and $R_0$ also depends on the geometric size of the
BH binary, $R_0\sim G(M_1+M_2) a/J_c^2$.  
Note that the product $\langle(\Delta
R)^2\rangle \bar f$ is independent of the number of stars in the galaxy
$N\sim M_{\rm galaxy}/m_*$.
The total mass flow $m_* \int \flux(E)dE$, however, scales as $N^{-1}$.

\section{Does the standard theory apply to MBH binaries?} 
\label{sec:issues}

The orbital separation $a$ of MBHs in a binary 
is much larger than the tidal disruption radius around an isolated MBH.
Using equation (\ref{eq:tidalradius}):
\bea
\frac{r_{\rm bin}}{r_{\rm tidal}}&\approx&
10^5 \times \eta^{-2/3} 
\left(\frac{M_\bullet}{10^8M_\odot}\right)^{-1/3} 
\left(\frac{m_*}{M_\odot}\right)^{1/3} \nonumber\\ & &\times
\left(\frac{r_*}{R_\odot}\right)^{-1} \left(\frac{a}{1\textrm{
    pc}}\right) .
\eea
Henceforth, $M_\bullet=M_1+M_2$ denotes the total mass in MBHs.
In the context of a loss cone associated with a binary, 
most of the galaxy is in the strongly diffusive regime 
and the boundary
layer is of negligible thickness, 
$R_0\sim R_\losscone$.  Figure \ref{fig:q}a shows the dependence of the
quantity $q$ on energy for a model described by the 
$\rho\sim r^{-2}(1+r^2)^{-2}$ stellar density profile and scaled to
the galaxy M32 with $0.1\%$ of the galaxy mass in the central MBH.  
Note that $q(E)\ll 1$ when the MBHs first form a
bound pair and remains $\ll 1$ even when the binary separation
shrinks to $0.01$ pc.  
Scaling to other galaxies, assuming a nuclear density profile of 
$\rho\sim r^{-2}$ and a fixed ratio of MBH mass to the galaxy mass,
we find:
\bea
\label{eq:qfit}
q(E)&\approx& 0.025 
\times\left(\frac{m_*}{M_\odot}\right)\left(\frac{M_\bullet}{3\times
  10^6M_\odot}\right)^{-1}\nonumber\\& &\times
\left(\frac{a}{GM_\bullet/8\sigma^2}\right)^{-1}
e^{-E/2\sigma^2}
\eea
with $\sigma$ the 1D stellar velocity dispersion.
Since $q(E)\sim \period(E)/T_{\rm
  relax}(E)$, 
the typical deflection
of a star due to the presence of other stars over one orbital period
is small compared with the size of the loss cone.
Therefore,
$f$ is nearly
uniform around the orbit of 
the star; the stellar system satisfies 
Liouville's theorem; and the orbit-averaged
Fokker-Planck equation \citep{lis77} is a valid
description of the loss cone dynamics.  
We will rely on this conclusion throughout
 this work.

Another aspect of the standard theory fares less well in the case of
  MBH binaries. 
  We have so far discussed the time-independent Fokker-Planck equation;
  strictly speaking, however, time-independent solutions do not exist as stars
  that diffuse into the loss cone are moved from one orbit to another,
or removed from the system. 
The flux of stars into the loss cone given by the standard theory
is shown in Figure \ref{fig:q}b. 
As the stellar orbital integrals diffuse, 
the potential changes accordingly, and the
  isotropic distribution $\bar f(E)$ adjusts to the changing
  potential.  \citet{Milosavljevic:01} demonstrated that the
  change of the overall density profile of the galaxy can amount to a
  destruction of the steep central cusp. 
  One may attempt to account for the changing galaxy profile by
  adjusting
  $\bar f(E)$ as dictated by the isotropic Jeans equation, while
  keeping the formal solution in equation (\ref{eq:soltimeind}) unchanged.  
Similarly, as the MBH binary hardens, the loss cone boundary
$R_\losscone$ decreases.  
One could try scaling the formal solution 
to accommodate a changing $R_\losscone$.

This strategy meets a serious 
obstacle when considering the
time required for the product of a galaxy
merger to converge to the quasi-steady state 
described in the previous section. 
The merger and the subsequent formation of a hard binary proceed on the local dynamical
time scale, which can be much shorter 
than the collisional relaxation time scale.
Therefore the distribution function $f(E,R)$
immediately following the formation of a hard binary can be far from 
the solution in equation (\ref{eq:soltimeind}).  
The mass in stars near the loss cone in excess of
the equilibrium solution can be 
orders of magnitute larger than the mass fed into the loss
  cone over a Hubble time assuming the equilibrium solution.  
Sudden draining of the loss cone during formation of the hard binary
produces steep phase space gradients 
that are closer to the step function:
\beq
f(E,R) \approx   
\cases{ \bar f(E), & $R>R_\losscone$ \cr
  0 , & $R<R_\losscone$ } .
\eeq
Since the collisional transport rate in phase space is proportional
to the gradient of $f$ with respect to $R$,
steep gradients imply an enhanced flux into the loss cone
and an accelerated evolution
toward the equilibrium form.

\begin{inlinefigure}
\begin{center}
\resizebox{0.8\textwidth}{!}{\includegraphics{figure1.epsi}}
\end{center}
\figcaption{\label{fig:q}
Characteristic quantities associated with loss cones in galaxies
  described by the \citet{jaf83} model,
  $\rho\sim r^{-2}(1+r^2)^{-2}$.
  Plots have been scaled to match the galaxy M32 with 
  a $3\times10^6 M_\odot$ central MBH.  The energy is expressed in
  terms of the central 1D velocity dispersion $\sigma$.  
  (a) The function $q(E)$. From bottom up,
  solid curves correspond to loss cones with capture radii of 1 pc, 0.1 pc,
  and 0.01 pc, respectively.  Note that $q(E)\ll 1$ in all cases, indicating
  that the loss cone associated with a binary MBH at the center of a galaxy 
  like
  M32 is always in the diffusive regime. 
  (b) The flux $\flux$ of
  stars into the loss cone assuming the standard result reviewed in
  \S~\ref{sec:reviewck}; the flux is expressed in MBH masses per unit
  $E/2\sigma^2$ per Myr.  
  (c) The time scale in Myr on which
  collisional relaxation can cause a significant change in the shape 
  of the stellar distribution function $f(E,R)$, assuming that the all stars
  have the same mass, $m_*=1M_\odot$.
  Convergence of
  arbitrary initial conditions to the standard loss cone solution (equation
  \ref{eq:soltimeind}) takes place on this time scale.}
\end{inlinefigure}

Figure \ref{fig:q}c shows that in low
mass ellipticals like M32, the time needed to achieve equilibrium 
depends on the energy, but still measures
well in excess of the Hubble time for the relevant range of energies.  
In intermedate-mass and massive
galaxies, the time is longer than the Hubble time regardless of the
orbital energy.
In \S~\ref{sec:timedep} we derive
equations describing the evolution of the stellar distribution
starting from arbitrary post-merger initial conditions and quantify
the discrepancies with the steady state solution.  We find that 
the steady-state solution is {\it never} 
a good description of a stellar system containing a binary MBH.

We have so far ignored the fate of the stars that are ejected by the
binary. Unlike the case of tidal disruption of stars by a MBH, a
  star ejected by a binary MBH via the gravitational slingshot
remains inside the galaxy and its host dark matter halo.
Ejected stars may therefore return to interact 
  with the binary for a second and subsequent times.  Every
  ejected star can, in principle, interact with the binary infinitely many
  times.  Its energy can increase {\it or} decrease at every
  encounter.  This process has been
  ignored in the published Fokker-Planck treatments of tidal disruption;
  nonetheless 
  it warrants consideration because of its capacity to shorten the lifetime of
  a MBH binary even after all of the stars inside the loss cone have been
  ejected at least once.  We discuss the efficiency of reejection in
 \S~\ref{sec:secondary}.

The standard treatment predicts that the decay time of MBH
  binaries scales with the inverse 
stellar mass $a/\dot a\sim m_*^{-1}\sim N$, where $N$
  is the number of stars in the galaxy.  Evidence that self-consistent 
$N$-body 
simulations reproduce this trend has so far been equivocal.  Simulations with
$M_\bullet/m_* \lesssim 10^3$ ($N\lesssim 10^5$), such as those 
reported in \citet{Milosavljevic:01},
appear to have taken place in
the ``pinhole'' regime, where the loss cone is almost full and the 
$N$-dependence is weak.  In simulations where the MBH binary was
allowed to wander in space, its Brownian motion was a likely source of
additional, spurious diffusion of stars in the phase space.  
Simulations with $M_\bullet/m_*\sim
10^4$ and a binary that does not wander are able to recover the trends
characteristic of the diffusive regime.
We discuss $N$-body simulations of MBH binary dynamics in 
\S~\ref{sec:nbody} and \S~\ref{sec:brownian}.   

\section{Time-Dependent Evolution of the Loss Cone}
\label{sec:timedep}

\subsection{The Loss Cone as an Initial Value Problem}
In this section we study the diffusion of stars into the loss cone of
a MBH binary following a merger of two galaxies.  
We assume that stars
are removed from the final galaxy the first time they are ejected by the
MBH binay; the continued role of these
stars is addressed in \S~\ref{sec:secondary}.
The loss cones of galaxies harboring MBHs are in the diffusive
regime.  A large fraction of stars scattering into the
geometric loss cone are kicked out by the binary in one orbital period.  This
suggests that $R=R_\losscone(E)$ can be thought of as an absorbing boundary
condition where the distribution function vanishes.  Since $R_\losscone(E)$
has no explicit dependence on the orbital phase, 
one can seek solutions of the
Fokker-Planck equation that have no explicit dependence on $r$.  The only
remaining dependence on $r$ enters through the diffusion coefficients in the
collision term of the Boltzmann equation and can be integrated out.  As a
result, one obtains the ``orbit averaged'' Fokker-Planck equation 
(Lightman \& Shapiro 1977).  

Instead of the distribution function $f$, we work in 
terms of the number density of stars in the $(E,R)$ plane, which is
related to the distribution function via:
\beq
\en (E,R,t)dE dR = 4\pi^2 \period(E,R) J_c(E)^2 f(E,R,t) dE dR .
\eeq 
The time-dependent Fokker-Planck equation including terms that
describe diffusion in the angular momentum direction reads:
\beq
\label{eq:fp}
\frac{\partial \en}{\partial t} =
-\frac{\partial}{\partial R}\left(\langle\Delta R\rangle \en\right) 
+ \frac{1}{2}\frac{\partial^2}{\partial R^2}
\left[\langle(\Delta R)^2\rangle \en\right] .
\eeq
Using a relation \citep{bil88} 
between the the first- and the second-order diffusion 
coefficients,
\beq
\langle\Delta R\rangle=\frac{1}{2}\frac{\partial}{\partial R} 
\langle\left(\Delta R\right)^2\rangle ,
\eeq
equation (\ref{eq:fp}) can be simplified to obtain:
\beq
\frac{\partial \en}{\partial t}=
\frac{1}{2}\frac{\partial}{\partial R}
\left[\langle\left(\Delta R\right)^2
\rangle\frac{\partial \en}{\partial R}\right] .
\eeq
Next we expand the diffusion coefficient $\langle(\Delta R)^2\rangle$ in the
limit $R\rightarrow 0$:
\beq
\frac{1}{2}\langle(\Delta R)^2\rangle = 
\left[\lim_{R\rightarrow 0}
\frac{\langle(\Delta R)^2\rangle}{2R}\right]   R + {\cal O} (R^2) ,
\eeq 
where the coefficient in brackets depends on energy and radius. We 
then average over one orbital period $\period^{-1} \oint dr/v_r$ so that
the Fokker-Planck equation ($R\ll 1$) becomes:
\beq
\label{eq:fpnr}
\frac{\partial \en}{\partial t} = 4\mu \frac{\partial }{\partial R}
\left(R \frac{\partial \en}{\partial R}\right) ,
\eeq
where
\beq
\mu(E)\equiv \frac{1}{4\period}  \oint 
\frac{dr}{v_r}
\lim_{R\rightarrow 0}
\frac{\langle(\Delta R)^2\rangle}{2R} .
\eeq
Note that this definition of $\mu(E)$ differs from that in \citet{mat99} by a factor of $1/4$.
We finally carry out a change of variables $R=\rootr^2$ to 
rewrite the equation as:
\beq
\label{eq:diffusion}
\frac{\partial \en}{\partial t} = 
\frac{\mu}{\rootr}\frac{\partial }{\partial \rootr}
\left(\rootr \frac{\partial \en}{\partial \rootr}\right) .
\eeq
This is the heat equation in cylindrical coordinates 
with radial variable $\rootr$ and diffusivity $\mu$.  At every energy $E$,
boundary conditions need to be specified at $0< \rootrlc<1$ and
$\rootr=1$, where $\rootrlc\equiv \sqrt{R_\losscone}$.  
The boundary condition at $\rootr=1$ is of the Neumann type: 
\beq
\label{eq:bdryouter}
\left. \frac{\partial \en}{\partial \rootr}\right|_{\rootr=1}=0 .
\eeq
The boundary condition at $\rootr=\rootrlc$ is a perfectly
absorbing, Dirichlet boundary condition:
\beq
\label{eq:bdryinner}
\en(E,\rootr)=0\ \ \ \ \ \ (\rootr \leq \rootrlc)
\eeq 
The geometrical meaning 
of the boundary conditions is illustrated in
Figure \ref{fig:initial}.

\begin{inlinefigure}
\begin{center}
\resizebox{0.8\textwidth}{!}{\includegraphics{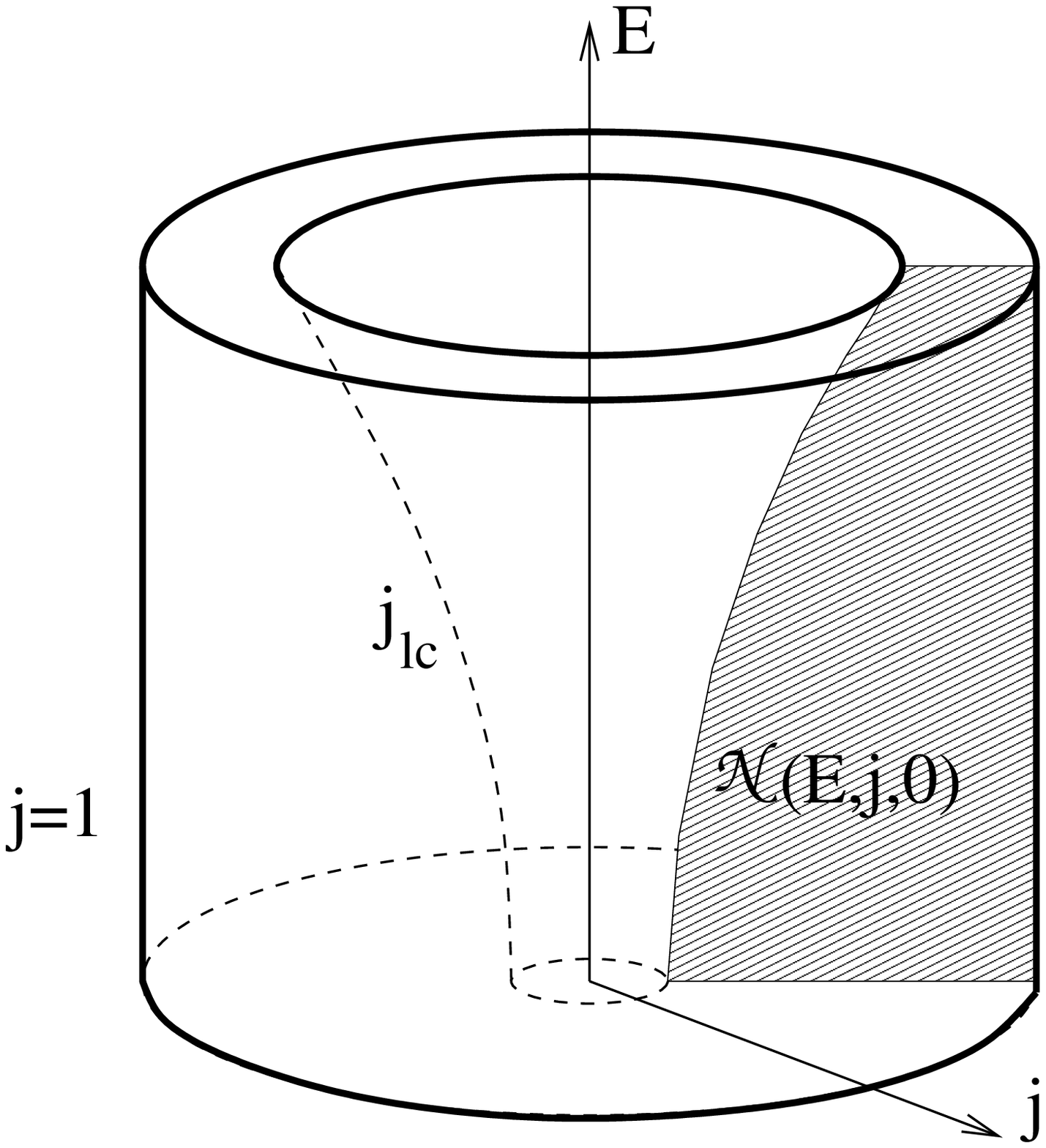}}
\end{center}
\figcaption{\label{fig:initial}
Schematic representation of the loss cone.  The orbital
  population function when the hard MBH binary forms,
  $\en(E,\rootr,t=0)$, is specified across the shaded surface
  (cylindrical symmetry is assumed).  For $t>0$, 
  the evolution of $\en$ is identical to the diffusion of heat into 
  the sink region $\rootr=\rootrlc$.  As the MBH shrinks,
  $\rootrlc$ moves inward.}
\end{inlinefigure}

The general solution of equation (\ref{eq:diffusion}) 
subject to boundary conditions 
(\ref{eq:bdryouter}) and (\ref{eq:bdryinner}) 
can be obtained by means of 
Fourier-Bessel synthesis (e.g.~\citealt{Ozisik:80}):
\bea
\label{eq:sink}
\en(E,\rootr,t)&=&
\frac{\pi^2}{2}\sum_{m=1}^\infty 
\frac{\left[\beta_m J_0(\beta_m\rootrlc)\right]^2}
{\left[J_0(\beta_m\rootrlc)\right]^2-\left[J_1(\beta_m)\right]^2}
\nonumber\\
& &\times
A(\beta_m;\rootr)
e^{-\mu\beta_m^2 t} 
\nonumber\\
& &\times
\int_\rootrlc^1 \rootr' A(\beta_m;\rootr')
\en(E,\rootr',0) d\rootr' ,
\eea
where $J_n$ and $Y_n$ ($n=0,1$) 
are Bessel functions of the first and second kind, $A(x;y)$ is a
combination of the Bessel functions defined via:
\beq
A(x;y)\equiv J_0(xy)Y_1(x)-J_1(x)Y_0(xy) ,
\eeq
while the $\beta_m$ are consecutive solutions of the equation: 
\beq
A(\beta_m;\rootrlc)=0 .
\eeq

We also evaluate the time-dependent version of the loss cone flux:
\bea
\label{eq:timedepflux}
\flux(E,t)dE&=&  -\frac{d}{dt}
\left[2\int^1_\rootrlc \en(E,\rootr,t) \rootr d\rootr\right]\nonumber\\
&=& 
-\pi^2\sum_{m=1}^\infty 
\frac{\mu{\beta_m}^3\rootrlc\left[J_0(\beta_m\rootrlc)\right]^2}
{\left[J_0(\beta_m\rootrlc)\right]^2-\left[J_1(\beta_m)\right]^2}
\nonumber\\
& &\times B(\beta_m;\rootr_\losscone)
e^{-\mu\beta_m^2 t}
\nonumber\\
& &\times
 \int_\rootrlc^1 \rootr' A(\beta_m;\rootr')
\en(E,\rootr',0) d\rootr' . 
\eea
where $B(x;y)$ is another combination of the Bessel functions
\beq
B(x;y)\equiv J_1(xy)Y_1(x)-J_1(x)Y_1(xy) .
\eeq

\subsection{History of the Loss Cone Profile}
In Figure \ref{fig:evolution}a we show the evolution of $\en(E,R,t)$ at one,
arbitrarily chosen $E$, assuming that
the loss cone is empty within
$R_\losscone$ and that $\en$ is a 
constant function of $R$ outside of $R_\losscone$ at the beginning
($t=0$).  Here we have assumed that the loss cone boundary is static, 
$R_\losscone\approx0.02$.  For comparison, we also show the equilibrium
solution of equation (\ref{eq:soltimeind}) which can be expressed:
\beq
\en_{\rm equi}(E,R,t)=
\frac{\ln(R/R_\losscone)}{\ln(1/R_\losscone)-1}\int_{R_\losscone}^{1}
\en(E,R,t) dR .
\eeq
The phase-space gradients 
$\partial\en/\partial R$ decay rapidly at first and then more gradually 
as they approach the equilibrium solution; 
the former converges to, but never becomes equal to
  the latter.  
It is evident that the total population $\int\en dR$ incurs 
a decrement of order unity before it has had time to reach 
the state of collisional equilibrium.  A decrease in $\en$
implies a decrease in the spatial density gradients in the
galaxy, or ``cusp destruction.''

\begin{inlinefigure}
\begin{center}
\resizebox{\textwidth}{!}{\includegraphics{figure3.epsi}}
\end{center}
\figcaption{\label{fig:evolution}
(a) Slices of the density $\en(E,R,t)$ at one,
  arbitrary $E$, recorded, from left to right, at $10^0$, $10^1$, $10^2$,
  $10^3$, and $10^4$ Myr (solid
  curve). Initially, $\en(E,R,t)=0$ for $R\leq R_\losscone$ and
  $\en(E,R,t)=\textrm{const}$ for $R>R_\losscone$.  We also show the
  equilibrium solution of equation (\ref{eq:soltimeind}) (dot-dashed
  curve).  
(b) The total number of stars consumed by the
  loss cone as a function of time (solid curve).  
  The scale has been set to galaxy M32
  modeled as a Jaffe model \citep{jaf83} with
  $M_\bullet=3\times10^6 M_\odot$ and an 
  initial separation between the MBHs of
  0.1 pc.}
\end{inlinefigure}

In Figure \ref{fig:evolution}b we present 
the total mass consumed by the binary, which is given by equation
(\ref{eq:eject}).    
The time-dependent and the equilibrium masses
are shown for comparison.
The equilibrium solution (dot-dashed curve) yields a loss cone
  flux smaller by at least a factor of two within a Hubble time
(Fig. \ref{fig:evolution}b).  
  The static loss cone boundary approximation employed here is valid 
  when $M_{\rm lost}\ll M_\bullet$ (\S~\ref{sec:approx}), 
or below the dashed line in the figure; we
  consider the general case in \S~\ref{sec:evolsemimajor}.
Our model here is that of a galaxy with a steep density profile such as M32.
The enhancement over the equilibrium fluxes is
the strongest in the galaxies with shallower central profiles or ``cores;'' 
there, however, 
$M_{\rm lost}$ remains smaller than $M_\bullet$ during a Hubble time assuming
perfect sphericity of the galactic potential.

The following remarks illustrate how the time dependence of the 
loss cone profile complicates the inferences that can be made about MBH 
binaries in real galaxies. 
Given $\rho(r)$, the Jeans equation can be solved for the isotropic
distribution function $f(E)$ and the orbital population $\en(E)\equiv
\int\en(E,R) dR$.
However from equation (\ref{eq:fpnr}), 
the flux of stars scattering into the loss cone is proportional
to the phase space gradient in $R$ at the boundary of the loss region,
or:
\beq
\flux(E,t)=4\mu R_\losscone 
\left.\frac{\partial\en}{\partial R}\right|_{R_\losscone} .
\eeq
If the nucleus is not collisionally relaxed,
the angular momentum dependence of $\en(E,R)$ is unknown.
Since the gradient $\partial\en/\partial R$ evolves in time 
(e.g.~equation \ref{eq:timedepflux}), the loss cone flux is 
likely to be a strong function of the age of the galaxy since
the most recent merger event, as well as of other details like
the mass ratio of the merger.

Attempts to estimate the separations of MBH binaries in presently 
observed galaxies (e.g.~\citealt{Yu:02a}) are therefore problematic.
As a first step, we would need to relate the total mass in stars
ejected by the MBH binary (following the initial emptying of the 
loss cone) to the evolution of its semi-major axis.  
The most commonly used form of this relation \citep{Quinlan:96}, 
derived via scattering experiments and confirmed
in $N$-body simulations in the case of 
equal-mass binaries \citep{Milosavljevic:01},
reads:
\bea
\label{eq:eject}
M_{\rm lost}(t)&\equiv& m_*\int_0^t\int\flux(E,t')dEdt'\nonumber\\
&\sim&\quinlanj M_\bullet\ln \frac{a(0)}{a(t)} ,
\eea
where, again, $M_\bullet=M_1+M_2$ is the binary mass, and 
$\quinlanj$ is a numerical factor approximately equal to one when the
binary is hard and $M_1\approx M_2$.
\citet{Milosavljevic:01} found that $\quinlanj\approx0.5$
correctly described the binary decay in their $N$-body simulations; in a
general case $\quinlanj$ is a function both of the hardness of the binary
and the MBH mass ratio $M_1/M_2$ (but see \S~\ref{sec:newj}).
Therefore the number of $e$-foldings in the binary separation is 
proportional to the mass ejected.   
Since the evolution of the binary separation depends on
the history of the loss cone flux from the binary formation onward, 
the assumption that the
equilibrium flux obtained from the {\it presently observed} 
$\rho(r)$ was the same at
all times in the past \citep{Yu:02a} is too conservative.  
Not only would the phase-space gradients $\partial\en/\partial R$ 
have been steeper in the past, but the density profile would have been 
steeper as well, providing significantly enhanced fluxes early in the 
life of the binary.  
Present-day MBH binary separations calculated assuming steady-state 
galaxies \citep{Yu:02a} should therefore be interpreted as upper limits.

\subsection{Approximations}
\label{sec:approx}
Several approximations have been made in deriving the solution in equations
(\ref{eq:sink}) and (\ref{eq:timedepflux}).  
These approximations relate to the assumption that the
isotropized distribution function $f(E)$ changes very little over the time
interval $t\in[0,t_{\rm max}]$ when the solution is expected to be valid.  We summarize
the approximations as follows.

1. The local crossing time is much shorter than the local relaxation time,
$\period(E)\ll T_{\rm relax}(E)$.  This is generally required of the
perturbative expansion on which the derivation of the Fokker-Planck is based.
Since we have calculated the time dependence explicitly, it is no longer
necessary that the system be old compared with the relaxation time.  It is
however imperative that the physical system modeled by equation 
(\ref{eq:diffusion}) have
reached dynamical equilibrium prior to the time the model can be considered
applicable. 

2. The shallowing of the overall potential due to the ejection of stars from
  the loss cone is small.  This guarantees that the diffusivity $\mu(E)$ also 
stays constant.  In the tidal disruption problem this is always the case, as
the consumed stellar mass is at most a small fraction of the MBH mass.
In the massive binary problem, the potential is {\it never} constant,
especially following mergers of galaxies with steep cusps when the binaries
impart significant damage to the merged cusp.  
Indeed, if the ejected mass
becomes comparable to the MBH mass---which is a likely outcome of
mergers of galaxies with steep density cusps---the potential near the MBH
will incur a relative change of order unity.  This notwithstanding, if the potential
changes slowly enough, adiabatic invariance suggests that the solution of
equation (\ref{eq:sink}) could be very close to the true solution modulo an
appropriate reparametrization of the energy dependence.

3. The geometric loss cone boundary $R_\losscone(E)=\rootrlc^2(E)$ 
changes little in time.  
Again, this is true for the tidal disruption loss cone,
but almost never true for the binary loss cone.  As the binary siphons energy
into stars that are captured and then ejected, the semi-major axis $a$ 
of the binary decreases.  Since $R_\losscone\sim a^2$, $\rootrlc$ shrinks
proportionally to the binary separation.  
The solution in equation (\ref{eq:sink})
will be valid only so long as $\Delta a\ll a$. 
The domain of validity of the static loss cone boundary approximation
can be estimated by considering $M_{\rm lost}(t)$.
Recall that the loss cone boundary scales with the binary separation,
$R_\losscone\sim a$.  
Therefore the approximation in which the geometric 
loss cone boundary is static is valid only if 
$M_{\rm lost}(t)\ll \quinlanj M_\bullet$.

\subsection{Evolution of the Semi-Major Axis}
\label{sec:evolsemimajor}
A self-consistent solution accounting for
the evolution of the loss-cone boundary in response to
the binary's orbital decay
can be constructed by iterating over
the appropriately chosen time steps.
The steps are chosen such that the binary separation changes little within
each step, $\Delta t\ll\quinlanj M_\bullet/m_* \int \flux(E,t) dE$.
At the end of each time step, one corrects 
the binary semi-major axis $a$ for the energy that the binary has
exchanged with the ejected stars as dictated by equation (\ref{eq:eject}).
One then recalculates the Fourier-Bessel
coefficients in equation (\ref{eq:sink}) using the new, advanced
$\en(E,\rootr,t+\Delta t_i)$ and the corrected loss cone boundary 
$\rootr_\losscone(E,t+\Delta t_i)$
and thus extends the time-dependent solution beyond $t+\Delta t_i$.
Iteration of this kind overcomes the requirement for a static background and
presents a computationally feasible numerical procedure for calculating loss
cone dynamics over cosmological times.  

We carried out this procedure
in a model of the galaxy M32 and present the results in Figure
\ref{fig:noneq}a--c.  The character of $a(t)$ does not vary with the choice
of the initial separation $a(0)$ as long as the initial width 
of loss cone is consistent with the separation chosen; 
here we have set the geometric loss cone boundary to equal
$R_\losscone(E) =GM_\bullet a/J_c^2(E)$.  
The initial separation used in the middle panel of Figure 
\ref{fig:noneq}, $a(0)=0.1$ pc, 
is the closest to what we expect to find in the merger of two
equal galaxies with $\sim 10^6 M_\odot$ MBHs.
Note that while $|da/dt|$ is bounded in the equilibrium solution 
(dashed line), 
in the time-dependent solution (solid line) $|da/dt|$ is
subtantially larger for small $t$ and diverges in the limit $t\rightarrow0$
due to the very large initial gradients $\partial\en/\partial R$.

\begin{inlinefigure}
\begin{center}
\resizebox{\textwidth}{!}{\includegraphics{figure4.epsi}}
\end{center}
\figcaption{\label{fig:noneq}
(a)-(c) Evolution of the semi-major axis in a singular isothermal sphere
  scaled to galaxy M32 with initial values of, from top to bottom, 1
  pc, 0.1 pc, and 0.01 pc.  Solid line is the evolution predicted by
  equation (\ref{eq:eject}); 
dashed line is the prediction of the equilibrium theory (\S~
  \ref{sec:reviewck}).  Scaling to other galaxies with $\rho\sim
  r^{-2}$ stellar density cusps can be achieved by linear
  re-parametrization of the time.  (d)-(f) Enhancement of the exact solution
  over the equilibrium solution (triangles).  The quantity $\Theta$ is defined
  in the text.  Power-law fits to the $\Theta(t)$ in the first 1 Gyr are also
  plotted (solid lines).}
\end{inlinefigure}

We can express the difference using
the quantity (denoting the equilibrium solution with $a_{\rm equi}$):
\beq
\Theta(t)\equiv \frac{a(0)-a(t)}{a(0)-a_{\rm equi}(t)} .
\eeq
In Figure \ref{fig:noneq}d--f we show the evolution of $\Theta(t)$; the
enhancement is well described by:
\bea
\label{eq:enhancetheta}
\Theta(t)\approx (10-15)\times 
\left(\frac{3\times10^6M_\odot}{M_\bullet}
\frac{m_*}{M_\odot}
\frac{t}{10\textrm{ Myr}}\right)^{-(0.33-0.39)} .
\eea
This relation is a good fit for 
$t\lesssim t_{\rm destr}$ where $t_{\rm destr}$ 
is the time scale on which the
removal of stars from the stellar cusp (the cusp destruction) 
starts affecting the pool of stars available for diffusion into the loss cone.
The time-scale can be identified with the diffusion of the stellar mass of
about ${\cal J}M_\bullet\sim 0.5M_\bullet$ into the loss cone, which in turn corresponds to the
change of the binary separation by the factor $e^{-1/2}\approx0.6$.  From
Figure \ref{fig:noneq}a-b, $a/a(0)=0.6$ at approximately at 1 Gyr for a
$3\times10^6M_\odot$ MBH. Scaling to an arbitrary binary mass gives:
\beq
t_{\rm destr}\sim\frac{M_\bullet}{3\times10^6M_\odot}
\left(\frac{m_*}{M_\odot}\right)^{-1}
\times1\textrm{ Gyr} .
\eeq
The difference between the two solutions decreases with time, to
roughly a factor of two after a Hubble time.

There is, however, a plausible scenario in which the mechanism 
discussed in this section could lead to a much greater enhancement 
in the binary's mean rate of decay.
Galactic nuclei may episodically return to a strongly non-equilibrium
state due to events like the capture of a dwarf galaxy, 
infall of a giant molecular cloud, or star formation from ambient gas.
A rejuvenated loss cone will once again result in strong
phase-space gradients as the binary rapidly ejects stars with
$R<R_{lc}$.
Then the the non-equilibrium enhancement averaged
over the entire lifetime of the binary would be higher than suggested by
Figure \ref{fig:noneq}.  For example, if the episodic replenishment of a MBH
binary loss cone in a galaxy like M32 occurs every 
10, 100, or 1,000 Myrs, equation
(\ref{eq:enhancetheta}) implies that the decay of the binary in that
galaxy will occur, respectively $\Theta\approx$ 
10, 5, or 3 times faster than what
the equilibrium theory would have predicted.

In conclusion: the equilibrium solution in equation 
(\ref{eq:soltimeind})---originally developed to describe the 
much smaller loss cones representing tidal disruption of stellar 
mass objects by massive ones---is never 
a good approximation to the structure of binary MBH nuclei.  
Instead, an epoch in which the flux of stars into the loss cone
is enhanced compared to the equilibrium solution, 
is succeeded by one in which the cusp is significantly modified due
to ejection of stars.
  
\section{Secondary Slingshot}
\label{sec:secondary}
\subsection{General Considerations}
\label{sec:reejectgen}
Until now we have ignored the possibility that the stars ejected by the
binary once are ejected again as they return to the nucleus on
radial orbits.  The stationary solution for 
the loss cone dynamics (\S~\ref{sec:reviewck}) 
and the time-dependent solution (\S~
\ref{sec:timedep}) are based on the ``sinkhole'' paradigm modelled
after the tidal disruption of stars by a MBH.  
In the tidal disruption picture
stars are destroyed as soon as they transgress the tidal disruption zone of the MBH.  
This is not true for MBH {\it binaries}.
The capture cross section of   
binaries is larger than that that of single MBHs 
by factors of $10^5-10^{11}$ and only a negligible
number of stars are disrupted; the vast majority simply receive a kick 
$(\Delta E,\Delta R)$ that transports them to another orbit. 
Following an event of slingshot, the 
final angular momentum is comparable to
the binary's orbital angular momentum, $J\sim \sqrt{GM_\bullet a}$.
If the binary orbit decays slower than the orbital period,
 $(d\ln a/dt)^{-1}\gg\period$, most stars inside the loss cone remain
inside the loss cone, encounter the binary at their next pericenter
passage, and are ``reejected.''
If the decrement in the binary separation is substantial, 
some stars that are ejected very near the loss cone boundary 
finish just outside boundary as the boundary shifts inward.  These
stars are lost at a rate proportional to the rate of decrease of the loss
cone angular momentum $R_\losscone\propto a$. 

In this section we explore the
possibility that the reejection can prolong the decay of a MBH binary
beyond the stalling point when the loss cone would have been 
depleted in the sinkhole paradigm. 
To simplify the forthcoming derivation, we consider an orbit-averaged
system in which  
probability for slingshot of star of energy $E$ located inside the
loss cone is uniform in time and totals to unity within one orbital period.  
Some stars can become quasi-permanently
captured by the MBH binary; in reality such captures are possible but
statistically negiligible.  Furthermore,
we assume that the orbital population function $\en(E,R,t)$ is
constant in $R$ inside the loss cone.  
We define the
loss-cone orbital population function: 
\beq
\en (E,t) \equiv \int_0^{R_\losscone} \en(E,R,t) dR .
\eeq
To derive the time-dependence of $\en(E,t)$, we consider the evolution over 
a small interval $\Delta t$.  The evolution has to account for the
redistribution of stars due to the kicks imparted by the MBH binary, as
well as for the gradual influx of stars due to the two-body relaxation.  
Therefore we have:
\beq
\label{eq:deltafinite}
\en(E,t+\Delta t) = \int \en(E',t) \zeta(E',E,\Delta t) dE' +
\flux(E,t) \Delta t .
\eeq
Here, $\zeta(E',E,\Delta t)$ is the transition probability between the energy 
$E'$ at time $t$ and the energy $E$ at time $t+\Delta t$, 
and $\flux(E,t)$ is the relaxation flux of stars 
as derived in \S~\ref{sec:reviewck} and
\S~\ref{sec:timedep}.  
Note that both depend implicitly on the binary's semi-major axis.
The transition probability can be understood as a combination
of the kicks from $E$ into some other energy, the
kicks from another energy $E'$ into $E$, and the attrition related to the
decrease in the loss cone size $R_\losscone$:  
\bea
\label{eq:zeta}
\zeta(E',E,\Delta t) &=&
\left[1-\frac{\Delta t}{\period(E')}\right] \delta(E-E') \nonumber\\
& &+
\frac{\Delta t}{\period(E')} 
\zeta_1 (E',E) 
 \nonumber\\
& &+ \left[\frac{R_\losscone(t+\Delta t)}{R_\losscone(t)} - 1\right] .
\eea
Here, $\zeta_1(E',E)$ is the probability 
that a star ejected from energy $E'$ will end up on an
orbit with energy $E$.  
The probability is normalized to unity $\int\zeta_1(E',E)dE'=1$.

Equation (\ref{eq:zeta}) is written under the assumption that the orbital
phases of stars at a given energy are randomly distributed.  In reality, this
may not be the case as stars retain the memory of their initial orbital
phases.   Variance in the ejection velocities will generate
dispersion in the distribution of stellar phases, which will 
destroy the phase coherence 
of the stars after they are ejected a few times.  We do not
believe that an initial non-uniform
distribution of the phases will have a significant
effect on the decay of black hole binaries.

Passing to the infinitesimal limit in equation (\ref{eq:deltafinite})
and using $d/dt\ln R_\losscone(t)=d/dt\ln a(t)$ 
we obtain an equation of motion for the orbital population of the loss cone:
\bea
\label{eq:lcpop}
\frac{\partial\en(E,t)}{\partial t} 
&=& -\frac{\en(E,t)}{\period(E)}
+ \int \frac{\en(E',t)}{\period(E')} \zeta_1(E',E)
dE' \nonumber\\
& &+ \en(E,t)\frac{d\ln a(t)}{dt} +
\flux(E,t) .
\eea
We proceed to derive an equation describing the evolution of
the total number of stars inside the loss cone:
\bea
\label{eq:evol_number}
\frac{d\en(t)}{dt}&\equiv& 
\frac{d}{dt} \int \en(E,t) dE\nonumber\\
&=&\en(t)\frac{d\ln a(t)}{dt}+\int\flux(E,t)dE ,
\eea
and of the total energy budget of the loss cone:
\bea
\label{eq:energy}
\frac{d {\mathcal E}(t)}{dt} &\equiv& 
\frac{d}{dt} \int \en(E,t) E dE \nonumber\\
&=& \int \frac{\en(E,t)}{\period(E)}
\left[-E + \int\zeta_1(E,E') E'dE'\right] dE \nonumber\\
& &+  \int \en(E,t) \frac{d\ln a(t)}{dt} EdE \nonumber\\
& &+ \int \flux(E,t) EdE .
\eea
The physical meaning of the factor in brackets in the last row of equation 
(\ref{eq:energy}) is immediately apparent---it is the average energy change
that a particle originally at energy $E$ experiences as a consequence of the
gravitational slingshot:
\beq
\langle\Delta E\rangle \equiv - E + \int\zeta_1(E,E')E'dE' .
\eeq

The energy gained in the first term of equation
(\ref{eq:energy}) has to be compensated by the change in the binary's binding
energy
\beq
\label{eq:rechard}
\frac{d}{dt} \left(\frac{G M_1 M_2}{2a}\right) = - m_*
 \int \frac{\en(E,t)}{\period(E)} \langle\Delta E\rangle dE .
\eeq
The hardening of a MBH binary coupled to an evolving 
stellar population inside the loss cone is 
described by equations
 (\ref{eq:lcpop}) and (\ref{eq:rechard}).  These equations are rendered in a
 form conducive to numerical integration.  

\subsection{Example: The Singular Isothermal Sphere}

To obtain a 
sense of how the secondary slingshot contributes to the decay of the
binary, we consider the following 
example: the galaxy is a singular isothermal
 sphere (SIS) with the density law:
\beq
\label{eq:sis}
\rho(r)=\frac{M}{4\pi r_0 r^2} ,
\eeq
and the potential:
\beq
\Phi(r)=-\frac{GM}{r_0}\ln\left(\frac{r}{r_0}\right) ,
\eeq
where $M$ and $r_0$ are constants.  Note that the potential due to this
 configuration diverges at $r\rightarrow\infty$; 
we have been free to fix $\Phi(r_0)=0$.  

The SIS is probably a better approximation to the
potential of a generic early-type 
galaxy than it may seem.  Indeed, while the luminous
matter (the stars) in a galaxy follows the SIS profile only within a
fraction of the effective radius, the {\it combined} stellar and CDM
contributions to the potential appear to approximate the SIS over and beyond
the extent of the luminous distribution. 
For simplicity we also assume that relaxation is absent in the model: 
$\flux(E,t)=0$.  The radial period inside the SIS potential is:
\beq
\period(E)=\oint \frac{dr}{v_r(E)} = \frac{\sqrt{\pi} r_0}{\sigma} 
e^{-E/2\sigma^2} ,
\eeq
where $\sigma\equiv\sqrt{GM/2r_0}$ is the velocity dispersion outside the
 radius of influence of the MBH.  Substituting this into equation
 (\ref{eq:rechard}) gives:
\beq
\label{eq:wperiod}
\frac{d}{dt} \left(\frac{G M_1 M_2}{2a}\right) = 
- \frac{m_*}{\period(0)}
 \int \en(E,t) e^{E/2\sigma^2}
\langle\Delta E\rangle dE .
\eeq
Naively it may appear that the integral in equation (\ref{eq:wperiod})
diverges.  This not the case because large (very bound) energies are
forbidden inside the loss cone, 
$\en(E,t)=0$ for $E>E_{\rm max}$, since the coexistence of a star
with large binding energy and a binary is very improbable.  
The distribution of stars inside the loss cone peaks at the energy
corresponding to the slingshot ejection velocity 
(see Figure \ref{fig:peak}):
\beq
\label{eq:peak}
E_{\rm eject} \sim \Phi_{\rm eject} - \frac{1}{2} \frac{G(M_1+M_2)}{a} ,
\eeq
where $\Phi_{\rm
  eject}$ is the fiducial potential energy of a star at the edge of
the MBH binary's sphere of influence
and $\sqrt{G(M_1+M_2)/a}$ is the orbital velocity the MBHs.  To be able
to solve equation (\ref{eq:wperiod}) analytically, we pretend that all
stars inside the loss cone have the same energy equal to $E_{\rm
  eject}$ at time $t$:
\bea
\en(E,t)\rightarrow \en(t) \delta(E-E_{\rm eject}) .
\eea
Therefore:
\beq
\label{eq:e_effective}
\frac{d}{dt} \left(\frac{G M_1 M_2}{2a}\right) \approx
- \frac{m_*}{P(0)}
 \en(t) 
\langle\Delta E\rangle_{E_{\rm eject}}e^{E_{\rm eject}/2\sigma^2}
\eeq
Equation (\ref{eq:e_effective}) can be simplified further after analyzing
the evolution of the product 
$\en(t) \langle\Delta E\rangle_{E_{\rm eject}}$.  
Equation (\ref{eq:evol_number}) implies that in the absence of
relaxation, the total number of stars inside the loss cone
decays proportionally to the binary separation, $\en(t)\propto
a(t)$. We further argue that $\langle\Delta E\rangle_{E_{\rm eject}}$
exhibits scaling with $a^{-1}(t)$, 
by considering two extreme cases:

\begin{inlinefigure}
\begin{center}
\resizebox{\textwidth}{!}{\includegraphics{figure5.epsi}}
\end{center}
\figcaption{\label{fig:peak}
Left panel, evolution of the distribution of stars initally inside the loss
cone $\en(E,t)$ in a simulation with $\en\approx18,000$ stars
inside the loss cone at $t=0$.  The distributions, from right to left, were 
recorded at logarithmically progressing times
$t=(0,1,2,4,8,16,32,64,128,256)\times 4\times
10^{-4} P(0)$, where $P(0)=\pi^{1/2}r_0\sigma^{-1}$ is the radial
crossing time at $r_0$.  The binary separation decayed by the factor
of $\sim 4.0$ during the same interval. The MBH binary mass $M_\bullet$ 
is $10^{-4}$ of the SIS mass $M$,
and the initial binary separation is $0.1 GM_\bullet \sigma^{-2}$,
consistent with the separations found in merger simulations 
\citep{Milosavljevic:01}.  The smooth curves were generated from 
$N$-body data via a maximum penalized likelihood technique.  Peak of
the distribution travels from larger to smaller binding energies
following the increase in the energy of ejection (equation
\ref{eq:peak}).
Note the late formation of the secondary hump at $E/(2\sigma^2)\sim3.0-8.0$, 
levelling at $\en\sim 1,400$; this hump consists of stars that had
originally been inside the loss cone, but then finished just
outside, out of reach to the MBH binary.  Right panel shows 
${\cal D}\equiv 10^3\partial\en(E,t)/\partial\ln t$ at 
$t=(1,2,4,8,16,32,64,128)\times 4\times 10^{-4} P(0)$.  The
zero-intercept indicates the energy at which the average exchange of
energy with the MBH binary vanishes, $\langle\Delta E\rangle=0$.}
\end{inlinefigure}

1. Just
after two MBH form a hard binary, the stars in the inner stellar cusp 
inside the loss cone
interact with the MBH binary for the first time and receive a
kick comparable to, if not larger than, than their kinetic energy:
\beq
\Delta E\gtrsim \frac{v^2}{2}, \ \ \ \ \ \ \ 
\left(\frac{v^2}{2}+\Delta E\right) \sim \frac{1}{2}\frac{G(M_1+M_2)}{a} .
\eeq
Thus $\langle\Delta E\rangle\propto a^{-1}(t)$ is indeed a good
approximation. 

2. When the binary is old and evolving slowly, one may wonder whether
a steady state ($\langle\Delta E\rangle=0$) can be 
achieved, in which the stars 
in the loss cone exchange no net energy with
the binary.\footnote{We thank C. Pryor for raising this question.}  
A steady state might be expected when 
the majority of stars inside the loss cone
have been ejected to large energies at which the energy exchange with the
binary is very inefficient. 
Assuming that we have indeed succeeded preparing such a steady state, 
we recognize that the second
moment $\langle(\Delta E)^2\rangle$ must be finite.  Defining
$\delta E\equiv \langle(\Delta E)^2\rangle^{1/2}>0$, we consider
the radial periods of 
stars ejected slower than the average ($E_1=\bar E+\delta E$) and
faster than the average ($E_2=\bar E-\delta E$).  Since the former
return to re-encounter the binary sooner than the latter,
$\period(E_1)<\period(E_2)$, at time $t+\period(\bar E)$ the
distribution of stars impinging on the binary will have 
acquired a finite first moment:
\beq
\bar E\rightarrow \bar E + 
\left(\delta E\right)^2\left. \frac{d\ln\period^{-1}}{dE}\right|_{\bar E} =
\bar E + \frac{\left(\delta E\right)^2}{2\sigma^2} .
\eeq
Thus the average effective energy of stars as experienced by the binary is
shifted by $(\delta E)^2/2\sigma^2$, and the binary will respond with
a decrease 
of the semi-major axis by $\Delta a=-a^2 (\delta E)^2
\en m_*/GM_1M_2\sigma^2$.  
This in turn will generate a shift in $E_{\rm eject}$ by the amount:
\bea
E_{\rm eject} &\rightarrow& E_{\rm eject}- \frac{G(M_1+M_2)}{2a^2}
\Delta a \nonumber\\
&=& E_{\rm eject} - \frac{\en m_*}{\mu}\frac{(\delta
  E)^2}{2\sigma^2} ,
\eea
where $\mu\equiv M_1M_2/(M_1+M_2)$ is the reduced mass.  Finally, the
true distribution of stellar energies shifts in response to the
incremental binary hardening by the amount
$\langle\Delta E\rangle=\Delta E_{\rm eject}$.  Recall that $\en\propto
a$.  From the scaling of the restricted three-body problem we know
that $\delta E\sim G(M_1+M_2)/2a$.  Therefore, again:
\beq
\langle\Delta E\rangle \sim \frac{\en(0) m_*}{\mu} \frac{a(t)}{a(0)} 
\frac{G^2(M_1+M_2)^2}{8\sigma^2 a^2(t)}
\propto a^{-1}(t) .
\eeq

We have shown that when the binary just becomes hard and evolves
rapidly, as well as when it is very hard and evolves very gradually,
the product $\en(t)\langle\Delta E\rangle\propto a^1a^{-1}\propto a^0$
is approximately constant (although the constant of proportionality
could be different in the above 
two regimes).  We proceed to integrate equation 
(\ref{eq:e_effective}) where we substitute $E_{\rm eject}$ from
equation (\ref{eq:peak}):
\beq
\label{eq:logsemi}
\frac{1}{a(t)} =
\frac{1}{a(0)}+
\frac{4\sigma^2}{G(M_1+M_2)}
\ln\left[1 + 
  \frac{m_* \en \langle\Delta E\rangle}{2\mu\sigma^2}
  \frac{t}{\period(E_0)} 
\right]
\eeq
for $t>0$,
where $a(0)$ is the initial separation and $E_0$ is the energy of
ejected stars at the outset:
\beq
E_0=\left.E_{\rm eject}\right|_{t=0} = \Phi_{\rm eject} - \frac{G(M_1+M_2)}{2a(0)} .
\eeq

Equation (\ref{eq:logsemi}) describes the evolution of a MBH binary in the SIS
potential ignoring relaxation. 
In the secondary slingshot mechanism, the inverse
semi-major axis and, hence, the binding energy, are {\it logarithmic} 
functions time.  
This result has been derived assuming
that relaxation-driven loss-cone diffusion was negligible.  
The logarithmic behaviour can be contrasted
with the effect of diffusion in which 
the binding energy grows linearly.
In a realistic galaxy both mechanisms are at work; the reejection
amplifies the effect of diffusion on the hardening rate of the MBH binary.

It is
encouraging that the logarithmic behavior seems to hold in the
simulations where
the interstellar interactions have been replaced by a smooth potential to
prevent relaxation.  Figure \ref{fig:recyclesemi} shows the evolution
of the inverse semi-major axis in a simulation with $\en\approx
18,000$ stars in the loss cone at $t=0$.   The observed rate
of decay $da^{-1}/d\ln t\approx 1.7\times 10^4$ comes close to the
prediction of equation (\ref{eq:logsemi}): 
$4\sigma^2/G(M_1+M_2)=2\times10^4$. 

\begin{inlinefigure}
\begin{center}
\resizebox{0.8\textwidth}{!}{\includegraphics{figure6.epsi}}
\end{center}
\figcaption{\label{fig:recyclesemi}
Decay of the inverse semi-major axis $1/a(t)$ in an $N$-body 
  simulation where
  the stellar potential has been replaced by a smooth component, thereby
  preventing relaxation (solid line).  
  The binary separation does not stall, but continues
  to decay because of the seconary slingshot (\S~
  \ref{sec:secondary}).
  The galaxy is a singular 
  isothermal sphere (equation (\ref{eq:sis})) with $M=r_0=1$.
  The time is expressed in the units of
  $P(0)=\pi^{1/2}r_0\sigma^{-1}$, which is the radial crossing time at $r_0$. 
  To compare this curve with the model presented in equation
  (\ref{eq:logsemi}), the hard-binary separation $a(0)$ has to be
  chosen; it will generally be different from $a$ at the beginning of
  the simulation. We show prediction of the model with
  $a(0)=5\times10^{-6}$ and $m_*\en\langle\Delta
  E\rangle/2\mu\sigma^2=2\times10^5$ (dashed line).}
\end{inlinefigure}

\subsection{Reejection in Galaxies}
\label{sec:reejectgal}
An upper limit on the effectiveness of the
mechanism presented here can be estimated as follows. 
In a merger of two galaxies, the MBHs form a hard binary when $a(0)\sim a_{\rm
  hard}=
G\mu/4\sigma^2$. 
The total mass of
stars inside the loss cone just after the hard binary forms is
$m_*\en(0)\sim 10.0\mu$.  We can also assume that $M_1=M_2$ and
$\langle\Delta E\rangle\sim \textrm{few}\times2\sigma^2$.  
In a Hubble time, $a^{-1}$ will have increased by the factor:
\beq
\label{eq:reejectgal}
\frac{a(0)}{a(t_{\rm Hubble})}\sim
1+0.25\ln \left[\textrm{few}\times 10 \times
\frac{t_{\rm Hubble}}{P(E_0)}\right] .
\eeq
For $P(E_0)=10^{3,5,7}$ years, we obtain $a(0)/a(t_{\rm Hubble})\approx
5$, $4$, and $3$, respectively.

This conclusion may be optimistic.  
In a circular binary, orbital velocities of the MBHs can be written as:
\beq
V_1=2\sigma\sqrt{\frac{a_{\rm hard}}{a}\frac{M_2}{M_1}}, \ \ \ \
V_2=2\sigma\sqrt{\frac{a_{\rm hard}}{a}\frac{M_1}{M_2}}
\eeq
In the isothermal sphere potential, 
the stars ejected with velocities $V_1\lesssim v\lesssim V_2$ 
will leave the binary's sphere
of influence $r_{\rm h}=GM_\bullet/2\sigma^2$ 
and travel to a radius $r_{\rm max}$:
\beq
\exp\left(\frac{M_2}{M_1}{a_{\rm hard}\over a}\right)
\lesssim
\frac{r_{\rm max}}{r_{\rm h}} \lesssim
\exp\left(\frac{M_1}{M_2}{a_{\rm hard}\over a}\right).
\label{eq:rmax}
\eeq
Realistically, reejection becomes ineffective if $r_{\rm max}$ is
too large: first because the isothermal sphere potential may
not extend indefinitely; and second because a star with large
$r_{\rm max}$ is easily perturbed from its nearly-radial orbit on 
the way in or out.
If we suppose that $r_{\rm max}\sim 100r_{\rm h}$ 
is an effective upper limit,
equation (\ref{eq:rmax}) suggests that $a_{\rm min}\approx 0.2 a_{\rm hard}$ 
is an effective lower limit for reejection when 
$M_1\approx M_2$.

However, we note that many stars are ejected with velocities
much less than the binary's orbital velocity (the velocity change 
varies between 0 and $\textrm{few}\times V_2$, and can also be negative, 
depending on the star's orbital parameters and the
binary's phase). 
This is particularly true when
$M_2\ll M_1$; the average velocity change of a star encountering the binary is
$\Delta v\ll V_2$, 
although every star inside the loss cone can
be pumped to an ejection at $\sim V_2$
after some number of encounters with the binary.  If $\Delta v\sim V_1$, which
is statistically favored when $M_2\ll M_1$, the lower limit on the semi-major
axis is weakened, $a_{\rm min}\ll 0.2 a_{\rm hard}$.
We observe that the transfer of energy 
from the binary to the stars is an
incremental, stochastic process involving multiple encounters; the formalism
developed
in this section is applicable whenever the form of
the galaxy's gravitational potential 
permits repeated stellar encounters with the MBH binary.

\subsection{A New Mass-Ejection Formula}
\label{sec:newj}

The reejection paradigm suggests a new and more general expression
for the relation between the ejected mass and the shrinkage factor of a binary.
The standard expression, equation (\ref{eq:eject}), was motivated by
scattering experiments in which stars are assumed lost
if they exit the binary's sphere of influence with
enough velocity to escape the binary.
In the case of a binary embedded in a galactic potential,
the critical quantity is the energy gained by a star between
the time it enters and exits the loss cone.
Equating this with the change in the binary's absolute binding energy,
we find:
\beq
\label{eq:newbalance}
\frac{G(M_1+M_2)\mu}{2}\left(\frac{1}{a_{\rm final}}-
\frac{1}{a_{\rm initial}}\right)
= M_{\rm lost}(\bar E_{\rm enter} - \bar E_{\rm exit}).
\eeq
In an SIS potential, the greatest contribution to the 
loss cone comes from stars near $r_{\rm h}$. Defining 
$(\bar E_{\rm enter} - \bar E_{\rm exit})\equiv\Delta\Phi$ and
passing to the infinitesimal
limit $a_{\rm initial}\rightarrow a_{\rm final}\equiv a$ 
in equation (\ref{eq:newbalance}), we obtain:
\beq
\label{eq:reejectlaw}
\frac{1}{(M_1+M_2)}
\frac{dM_{\rm lost}}{d\ln a^{-1}}=\frac{1}{(\Delta\Phi/2\sigma^2)(a/a_{\rm hard})}
\eeq
Comparing this relation to equation (\ref{eq:eject}) we identify
the effective value of the mass-ejection parameter $\quinlanj$:
\beq
\label{eq:reejectj}
\quinlanj\sim \frac{1}{(\Delta\Phi/2\sigma^2)(a/a_{\rm hard})}.
\eeq 
To shrink the binary by one $e$-folding,
a stellar mass of $\quinlanj M_\bullet$ 
must be transported from an energy marginally bound to the black hole, 
to the galactic escape velocity.

\subsection{Reejection in Other Geometries}
Reejection would occur differently in nonspherical
(axisymmetric or triaxial) potentials, since angular
momentum would not be conserved and ejected stars could miss
the binary on their second passage.  However there will
generally exist a subset of orbits defined by a maximum
pericenter distance that is $\lesssim a$ and stars on such
orbits will interact with the binary once per crossing
time, just as in the spherical case. As an example, consider
centrophilic (chaotic) orbits in a triaxial nucleus
\citep{Poon:02}. There is one such orbit per energy
(in a time-averaged sense), and the cumulative distribution
of pericenter distances for a single orbit is found to be
linear in
$r_{\rm p}$ up to a maximum value, $r_{\rm p,max}(E)$ 
\citep{Merritt:03}.
Thus the star passes within a distance $r_{\rm p,max}$ of the
center each crossing time.
\citep{Merritt:03} evaluated $r_{\rm p,max}(E)$ for chaotic
orbits in the self-consistent triaxial models of 
\citet{Poon:02}, which have 
density profiles $\rho\sim r^{-2}$, and found:
\beq
r_{\rm p,max}(E)\approx 0.3 r_{\rm h}e^{\left(\Phi_{\rm h}-E\right)/2\sigma^2}
\eeq
with $\Phi_{\rm h}\equiv\Phi(r_{\rm h})$ and 
$r_{\rm h}=GM_\bullet/2\sigma^2$.
The probability of a star passing within
$a$ during a single traversal of the galaxy is 
$\sim a/r_{\rm p,max}$ for $a<r_{\rm p,max}$ and 
one for $a>r_{\rm p,max}$.
It follows that all stars on chaotic orbits with 
$E\gtrsim \Phi_{\rm h}$ pass inside of $a_{\rm hard}$ on each orbit, 
and a substantial fraction of
stars would continue to engage in reejection
even after the binary had shrunk by a factor of several below
$a_{\rm hard}$.
Hence reejection in the triaxial geometry might occur in roughly the
same way we have described for spherical systems.  The situation
would be more complex in axisymmetric (nonspherical) galaxies but there
would always exist some subset of orbits defined by
$r_{\rm p,max}<a$ particularly if the deviations from spherical symmetry
were modest.

\section{$N$-Body Simulations and Real Galaxies}
\label{sec:nbody}
In this section, we ask whether the long-term evolution of
MBH binaries seen in $N$-body simulations is consistent with
the theory derived above, and whether the simulations are capable
of reproducing the evolution of binaries in real galaxies.

Does the standard theory (\S~\ref{sec:reviewck})
reproduce the correct time-dependence for the evolution 
of the binary semi-major axis?  To solve for $a(t)$, we start 
from the differential form of equation
(\ref{eq:eject}):
\beq
\frac{d}{dt}\ln\left(\frac{1}{a}\right) 
= \frac{1}{\quinlanj M_\bullet}\frac{dM_{\rm lost}}{dt}
=\frac{m_*}{\quinlanj M_\bullet} \int \flux (E,a,t) dE .
\eeq
Note that both $\quinlanj$ and $\flux$ are functions of $a$.
In the equilibrium theory (equation \ref{eq:fluxtimeind}), 
$\flux\propto\ln^{-1}(a_1/a)$ 
where $a_1=J_c^2(E)/GM_\bullet e$ is independent of $a$.
Near the radius of
influence of the MBH where the peak loss cone flux originates,
$a_1\sim GM_\bullet/2\sigma^2$.  
Figure 5a of \citet{Quinlan:96} implies that for hard binaries,
\beq
\label{eq:quinlanj}
\quinlanj\approx0.1\times\ln\left(\frac{4 a_{\rm hard}}{a}\right),
\eeq 
where as before $a_{\rm hard}=G\mu/4\sigma^2$.
Therefore the evolution of the semi-major
axis in the standard model has the approximate form:
\beq
\frac{d}{dt}\ln\left(\frac{1}{a}\right)\approx
\frac{\Gamma/N_\bullet}{\ln(a_1/a)\ln(a_2/a)} .
\eeq
Here, $\Gamma/N_\bullet$ is a constant proportional to the 
angular momentum diffusion coefficient.  
We have also explicitly factored out the $N$-dependence
of this coefficient via $N_\bullet\equiv M_\bullet/m_*$,
the number of stars whose mass equals that of the MBH.
This equation can be easily solved
assuming $a_1\sim a_2\sim 4a_{\rm hard}$,
valid for approximately equal-mass MBHs ($M_\bullet=4\mu$),
yielding:
\beq
\label{eq:rootthree}
\frac{a_{\rm hard}}{a(t)} = \frac{1}{4}e^{(3\Gamma t/N_\bullet)^{1/3}} .
\eeq 
For $1<\Gamma t/N_\bullet<10$, the solution
(\ref{eq:rootthree}) is accurately approximated by the linear function:
\beq
\label{eq:standardn}
\frac{a_{\rm hard}}{a(t)} \approx \frac{1}{2}
\left(1+\frac{\Gamma t}{N_\bullet}\right)\ \ \ \ \ \ (1<\Gamma t/N_\bullet<10).
\eeq
For larger values of $\Gamma t/N_\bullet$, 
we are in the regime where the density
profile of the galaxy incurs significant damage and many assumptions of the
standard model do not apply.

In $N$-body experiments with $N\sim10^5$ and $N_\bullet\lesssim 10^3$ 
\citep{Milosavljevic:01} 
we observed a linear growth of $1/a(t)$.
However,  the binary decay in our simulations
went beyond the range $a(0)/a(t)\sim 1$ where the assumptions inherent in
the standard model are satisfied.  
Additional (unpublished) simulations with $N\sim10^6$ 
also produced a linear growth of $1/a(t)$.

Strikingly, however, these simulations did not yield the linear 
dependence of $1/a(t)$ on $N_\bullet$ predicted
by equation (\ref{eq:standardn}) (see \citet{Milosavljevic:01}, Fig. 8b).
Fearing that the Brownian motion of the MBH binary might have 
contributed to the loss-cone diffusion, 
we repeated the simulations in a symmetric mode that eliminated
the binary's Brownian 
motion.\footnote{Brownian motion will be identically nil 
in mergers of equal galaxies that are symmetric with respect to reflection 
$({\bf x},{\bf v})\rightarrow (-{\bf x},-{\bf v})$.}
The repeated experiments did yield a decrease in the
hardening rate with $N_\bullet$, but it was still much weaker than
linear. 
Similar behavior had been reported by \citet{mak97} 
whose simulated MBHs were more
massive relative to the galaxy than 
ours. Makino observed that the decay timescale was sub-linear in the number
of bodies, 
$d a^{-1}/dt\propto {N_\bullet}^{-1/3}$, which is still strongly at
odds with the prediction of equation (\ref{eq:standardn}).

The explanation for this discrepancy can be found in
Figure \ref{fig:qsim}, showing the quantity $q(E)$ in these simulations.
Recall that $q(E)\ll1$ means that the loss cone is almost empty (``diffusion
regime'') while $q(E)\gg1$ implies a nearly-full loss cone
(``pinhole regime'').  
The figure shows that the simulations of \citet{Milosavljevic:01} 
took place almost entirely in the pinhole regime -- the loss cone
was nearly full at all times.
In \S~\ref{sec:issues} we argued that the MBH binary loss cones in 
real galaxies are in the diffusive regime throughout the relevant 
range of energies. 
In the full loss cone regime, the flux of stars into the loss cone 
is given by the number of stars inside the loss
cone boundary assuming no depletion, 
divided by the radial period on which these stars arrive to 
within the binary's sphere of influence, or:
\beq
\flux^{({\rm pinhole})} 
(E)dE \sim \frac{\en(E) R_\losscone(E)}{\period(E)} dE
\eeq
which is valid for $E<E_{\rm crit}\approx 9\sigma^2$ 
(see Fig. \ref{fig:qsim}), $q(E_{\rm crit})\equiv 1$. 
The mass flow into the maw of the
binary $m_*\int\flux^{({\rm pinhole})}(E)dE$
is independent of $N_\bullet$ and always much larger than in 
the diffusive case.  
Ignoring reejection and noting that
$R_\losscone\propto a$,
we find that the semi-major axis is predicted to be 
an exactly linear function of the time:
\beq
\frac{1}{a(t)}=\frac{1}{a(0)}+t
\int_0^{E_{\rm crit}}\frac{Gm_*\en(E)dE}{\quinlanj\period(E) J_c^2(E)},
\eeq
as observed in the $N$-body simulations \citep{mak97,Milosavljevic:01}.

\begin{inlinefigure}
\begin{center}
\resizebox{0.8\textwidth}{!}{\includegraphics{figure7.epsi}}
\end{center}
\figcaption{\label{fig:qsim}
The quantity $q(E)$ in the simulated merger remnant 
  {\sf A2} of \citet{Milosavljevic:01} 
  with 131,072 bodies before truncation (solid
  line).  Also shown is $q(E)$ in an equivalent run with $10^6$ and no
  Brownian motion (dot-dashed line).  The energy is expressed in
  terms of the central linear velocity dispersion $\sigma$.  Note that in the
  merger remnant $q(E)\gg1$ even at energies as large as $8\sigma^2$, implying
  that the loss cone is in the pinhole regime.  This is the opposite of what is
  found in real galaxies like M32 that are entirely in the diffusive regime
  (see Fig.~\ref{fig:q}) and explains why $N$-dependence in the simulations
  was weak.}
\end{inlinefigure}

What value of $N$ is required in order that a numerical simulation
be in the appropriate, diffusive loss-cone regime?
The scaling of $E_{\rm crit}$ with $N$ can be deduced from
equation (\ref{eq:qfit}).
Assuming a density profile $\rho\sim r^{-2}$ and  
a potential of the form $2\sigma^2\ln(r/r_0)$ such that \
$r_0=10^3GM_\bullet/2\sigma^2$ we find:
\beq
\label{eq:ecrit}
E_{\rm crit}\approx 2\sigma^2 \ln \left(\frac{7.5\times10^4}{N_\bullet} 
\frac{GM_\bullet/8\sigma^2}{a}\right).
\eeq
The transition from a pinhole-like loss cone to a diffusive loss cone
occurs when $N_\bullet\sim 10^4-10^5$.  Since a typical MBH contains
0.1\% of its host galaxy's mass \citep{mef01a}, and thus $N\sim
10^3N_\bullet$, an $N$-body simulation would have to contain
$10^{4-5}\times10^3=10^{7-8}$ bodies to reproduce the correct,
diffusive behavior of a real galaxy.  
This requirement is a severe one for direct-summation $N$-body codes,
which rarely exceed particle numbers of $\sim 10^6$ even on
parallel hardware (e.g.~\citealt{Dorband:03}).
One route might be to couple the special purpose GRAPE
hardware\footnote{http://grape.astron.s.u-tokyo.ac.jp/grape/},
which is limited to $N\lesssim 10^6$, to algorithms that
can handle large particle numbers by swapping with a fast front end.

Finally, we can ask how reejection would affect the predicted 
time-dependence of the semi-major axis in $N$-body simulations,
assuming a loss cone always full at energies $E<E_{\rm crit}$.
For simplicity we assume that the Brownian motion has been suppressed;
the role of Brownian motion at enhancing binary decay is discussed
in detail in the next section.
Return and reejection of stars is a function of the 
depth of the potential well in which the model MBH binary has been placed.
The appropriate mass-ejection coefficient $\quinlanj$ 
is the one derived above, equation (\ref{eq:reejectj}),
which takes into account the total energy exchange with a star
before it finally escapes from the galaxy.
Note that $\Delta\Phi$---the potential barrier
between the energy at which stars are fed into the loss cone, and the energy
at which they escape the galaxy---is a function of loss cone entrance 
energy $E$. Then:
\bea
\frac{d}{dt}\ln\left(\frac{1}{a}\right)
&=&\frac{m_*}{M_\bullet}\int_0^{E_{\rm crit}}\frac{\flux^{({\rm pinhole})}(E)}{\quinlanj(E)}dE\nonumber\\
&=& \frac{{\cal D}}{2} 
\left(\frac{a}{a_{\rm hard}}\right)^2 ,
\eea
where, ignoring the implicit dependence of $E_{\rm crit}$ on $a$ (equation \ref{eq:ecrit}), 
${\cal D}$ is a constant:
\beq
{\cal D}=
\frac{G^2m_* \mu}{4\sigma^4}
\int_0^{E_{\rm crit}}\frac{\en(E)\Delta\Phi(E)}{\period(E)J_c^2(E)}dE .
\eeq
The integrand of ${\cal D}$ is independent of $N_\bullet$, 
however $E_{\rm crit}$
depends weakly on $N_\bullet$.  Integration gives:
\beq
\label{eq:semisqrt}
\frac{a_{\rm hard}}{a(t)}=
\sqrt{1+{\cal D}t} .
\eeq
This result and expression (\ref{eq:logsemi}) were both derived taking
reejection into account; their difference stems from the assumptions made
about the rate at which the loss cone is being refilled.  In equation
(\ref{eq:logsemi}), the refilling is absent and the binary decays
by reejection only.  In equation 
(\ref{eq:semisqrt}), the refilling is at its maximum
(as expected in $N$-body simulations) and
the reejection serves to amplify the effect of refilling 
on the binary's decay, converting a $\log t$ 
dependence to a $t^{1/2}$ dependence.

\section{Brownian Motion}
\label{sec:brownian}
Another potentially important source of flux 
into the loss cones of simulations in \citet{Milosavljevic:01} 
was the Brownian motion of the MBH binary.  Single and binary MBHs alike
acquire and maintain a random, quasi-periodic drift 
coming from the discrete encounters with 
stars at the center of the galaxy.  The drift can be understood
in terms of the equipartition of kinetic energy between the MBH(s) and
the ambient stars.
The Brownian motion was recently studied by \citet{mer01}, 
\citet{Chatterjee:02} and \citet{Dorband:03}.  
These studies concluded that the rms excursions
of the MBH's possition and velocity are given by (ignoring factors of
order unity):
\beq
\label{eq:brown}
\langle x^2 \rangle \sim\frac{m_*}{M_\bullet}  {r_{\rm core}}^2 ,
\ \ \ \ \ 
\langle \dot x^2\rangle \sim\frac{m_*}{M_\bullet} \sigma^2 ,
\eeq 
where $r_{\rm core}$ is the ``core radius'' of the stellar
distribution within which the stellar density approaches a finite,
central value.  In many real galaxies, however, the core radius is not
well-defined as the density increases near the MBH as an approximate
power-law, $\rho\sim r^{-\gamma}$ with $0\lesssim\gamma\lesssim 2$. 

$N$-body simulations \citep{Milosavljevic:01} reveal
that the MBH binary remained centered on the stellar
density cusp as it executes the Brownian motion; the binary ``carries
the cusp with it.''  The cusp endows the MBHs with a larger
effective dynamical mass.  We distinguish 
the stars closely following the MBH (``satellites''), 
from the stars that orbit within the stellar cusp and come close to 
the MBH during a fraction of their orbital period (``visitors'').  
The satellites respond to the binary's movement
and follow the binary as it wanders; their angular momentum with respect to
the binary remains approximately constant.  The visitors, however, 
stay embedded in the static galactic potential 
and experience the binary as moving target.  
As the binary moves, the visitors drift along nearly constant-energy 
trajectories in the $E$ -- $R$ plane that can penetrate
the loss cone boundary.  The drift can contribute significantly to the
loss cone refilling flux at energy $E$ if $\Delta R_{\rm brown}\gtrsim
R_\losscone$, where $\Delta R_{\rm brown}\sim GM_\bullet r_{\rm
brown}/J_c^2(E)$.

The meaning of the ``core radius'' in galaxies without a core in the density
profile can be understood in the light of the satellite-visitor dichotomy.
The satellites drift along with the MBH binary in the background of the
visitors and other stars not interacting with the binary.  The satellites, with
average velocities larger than the ambient stellar velocity dispersion, form a
bound subsystem in combination with the MBHs; the subsystem executes the 
Brownian motion in the field of other stars.  We can therefore think of the
background potential in which the binary-satellite subsystem moves as the
stellar potential with the contribution of the satellites {\it subtracted}.
The characteristic radial scale 
separating the satellites from other stars is the
the dynamical radius of influence of the black holes; therefore the effective
value for the ``core radius'' in coreless density profiles such as $\rho\sim
r^{-2}$ reads:
\beq
r_{\rm core} \sim \frac{GM_\bullet}{\sigma^2} .
\eeq

Assume the binary executes a small acceleration $\accel$, which could be
the associated with the Brownian motion or some other astrophysical
process.  We think of $\accel$ as a small perturbation to an otherwise
orbital integral-preserving motion of a test star.  
In the presence of Brownian motion, the acceleration can
be expressed using the quantities presented in equation
(\ref{eq:brown}):
\beq
\label{eq:accelbrown}
\langle\accel^2\rangle\sim \frac{\langle\dot x^2\rangle^2}{\langle x^2\rangle}
\sim \frac{m_*}{M_\bullet}\frac{\sigma^4}{{r_{\rm core}}^2} 
\sim \frac{m_* \sigma^8}{G^2 M_\bullet^3}.
\eeq
Now let ${\bf r}$ and ${\bf J}$ denote, respectively, 
the position and the angular momentum of the star {\it relative} to the
binary.  Then:
\beq
\frac{dJ^2}{dt}= 
2{\bf J}\cdot\frac{d}{dt}\left({\bf r}\times {\bf 
v}\right) 
= - 2{\bf J}\cdot({\bf r}\times \accel) .
\eeq
The total change in $J^2$ over one orbital period equals:
\beq
\Delta J^2=-2\int_0^{\period} {\bf J}\cdot ({\bf r}\times\accel) dt .
\eeq
Note that $\Delta J^2(E,{\bf J},\accel)
=\Delta J^2(E,J,|\accel|,\hat\Omega)$ is a function of the 
angles $\hat\Omega$ defining the inclination of the unperturbed orbit relative to
the acceleration vector.  
To calculate the second-order Fokker-Planck diffusion coefficient for the
diffusion in the $R=J^2/J_c^2(E)$ (see \S~\ref{sec:reviewck}), we square,
divide by the orbital period, and
average over $|\accel|$ and $\hat\Omega$:
\beq
\langle(\Delta R)^2\rangle
= \frac{
\left\langle\left[\Delta J^2(E,J,|\accel|,\hat\Omega)\right]^2
\right\rangle_{|\accel|,\hat\Omega}}{\period(E)J_c^4(E)}
\eeq
The diffusion time scale near the loss cone boundary at 
$J^2=J_\losscone^2(E)\sim GM_\bullet r_{\rm bin}$, where $r_{\rm bin}$
is the radius from the center of the binary within which gravitational
slingshot can occur (see \S~\ref{sec:reviewck}),
will then be:
\bea
\label{eq:tbrowngen}
t_{\rm brown}&=&
\frac{{R_\losscone}^2}{\langle(\Delta R)^2\rangle_\losscone}\nonumber\\
&\sim&
\frac{G^2 M_\bullet^2 r_{\rm bin}^2\period(E)}
{\left\langle\left[\Delta J^2(E,J_\losscone,|\accel|,\hat\Omega)\right]^2
\right\rangle_{|\accel|,\hat\Omega}}
\eea

To estimate $t_{\rm brown}$ we ignore the stellar self-gravity; the
test star is then orbiting in the approximately-Keplerian 
potential of the MBH binary.  
Consider an elliptical orbit with the closest approach to the
binary at $r_->r_{\rm bin}$.  
Let $\varphi$ denote 
the angle between $\accel$ and ${\bf J}$ (see Fig. \ref{fig:kepler}).
Since $dt=(r^2/J)d\theta$ 
where $\theta$ is the
azimuthal angle in the orbital plane,
$\Delta J^2$ becomes:
\beq
\label{eq:deltaell}
\Delta J^2_{\rm Kepler} = 
- 2|\accel|\sin\varphi \int_0^{2\pi} r^3\sin(\theta) d\theta .
\eeq
The perturbation in
the orbit is assumed linear in $\accel$; therefore it is sufficient to
substitute in equation (\ref{eq:deltaell}) the Keplerian orbit:
\beq
r(\theta)=\frac{GM_\bullet}{2E}\frac{(1-e^2)}{1+e\cos(\theta-\psi)} ,
\eeq
where $\psi$ is the inclination of the orbit with respect to the
projection of $\accel$ on the orbital plane.  Integration yields:
\beq
\label{eq:jkepler}
\Delta J^2_{\rm Kepler}=6 \pi |\accel| \left(\frac{GM_\bullet}{2E}\right)^3
e\sqrt{1-e^2}
\sin\varphi\sin\psi .
\eeq
Note that in Kepler's potential, $e^2=1-R$;
$J_c(E)=GM_\bullet/\sqrt{2E}$; and 
$\period(E)=2\pi GM_\bullet (2E)^{-3/2}$.  Therefore, 
knowing that $R_\losscone\ll 1$ at the loss cone
boundary, we have: 
\beq
\label{eq:ekepler}
e^2(1-e^2)_\losscone \approx R_\losscone \sim
r_{\rm bin}\left(\frac{GM_\bullet}{2E}\right)^{-1} . 
\eeq
Substituting equations (\ref{eq:jkepler}) 
and (\ref{eq:ekepler}) in equation 
(\ref{eq:tbrowngen}) gives the time scale 
on which Brownian motion refills the loss cone:
\beq
\label{eq:timebrown}
t_{\rm brown,Kepler}
\approx
\frac{(GM_\bullet)^{3/2}r_{\rm bin}}
{6\pi\accel^2}\left(\frac{GM_\bullet}{2E}\right)^{-7/2} .
\eeq
The time scale was derived ignoring the contribution to the mean potential
from the stars in the galaxy.  If the galactic potential had been taken into
account and the total potential had not been
Keplerian, the functional dependence
on $E$ in the time scale (\ref{eq:timebrown}) would have had a different form.
The scaling in the other parameters ($\accel$, $r_{\rm bin}$, $M_\bullet$),
however, will be the same in the generic case.  Substituting the expression
for the acceleration from equation (\ref{eq:accelbrown}) yields:
\bea
t_{\rm brown}&\sim& 400\textrm{ Myr}\times 
\frac{r_{\rm bin}}{a_{\rm hard}}
\left(\frac{m_*}{M_\odot}\right)^{-1}\nonumber\\ & &\times
\left(\frac{M_\bullet}{10^6M_\odot}\right)^{2-3/\alpha} 
e^{14/\alpha}  K(x) ,
\eea
where $a_{\rm hard}\equiv GM_\bullet/8\sigma^2$ 
is the hard binary separation, $\alpha\equiv d\ln
M_\bullet/d\ln\sigma$ is the slope of 
the relation between the black hole mass and
the velocity dispersion, and $K(x)$ is a 
dimensionless factor 
encapsulating the dependence on the energy 
$E\equiv2\sigma^2x$.  Using $\alpha\approx4.5$ \citep{fem00} and 
$r_{\rm bin}/a_{\rm hard}\sim 0.2$ in a galaxy like M32, 
we obtain $t_{\rm brown}\sim 5$ Gyr at
energies corresponding to the radius of influence of the black hole where
$K(x)\approx 1$.  
The total mass inside the loss cone at these energies is small; thus the
Brownian motion-driven diffusion of stars into the loss cone does not appear
to be efficient in galaxies.  

\begin{inlinefigure}
\begin{center}
\resizebox{\textwidth}{!}{\includegraphics{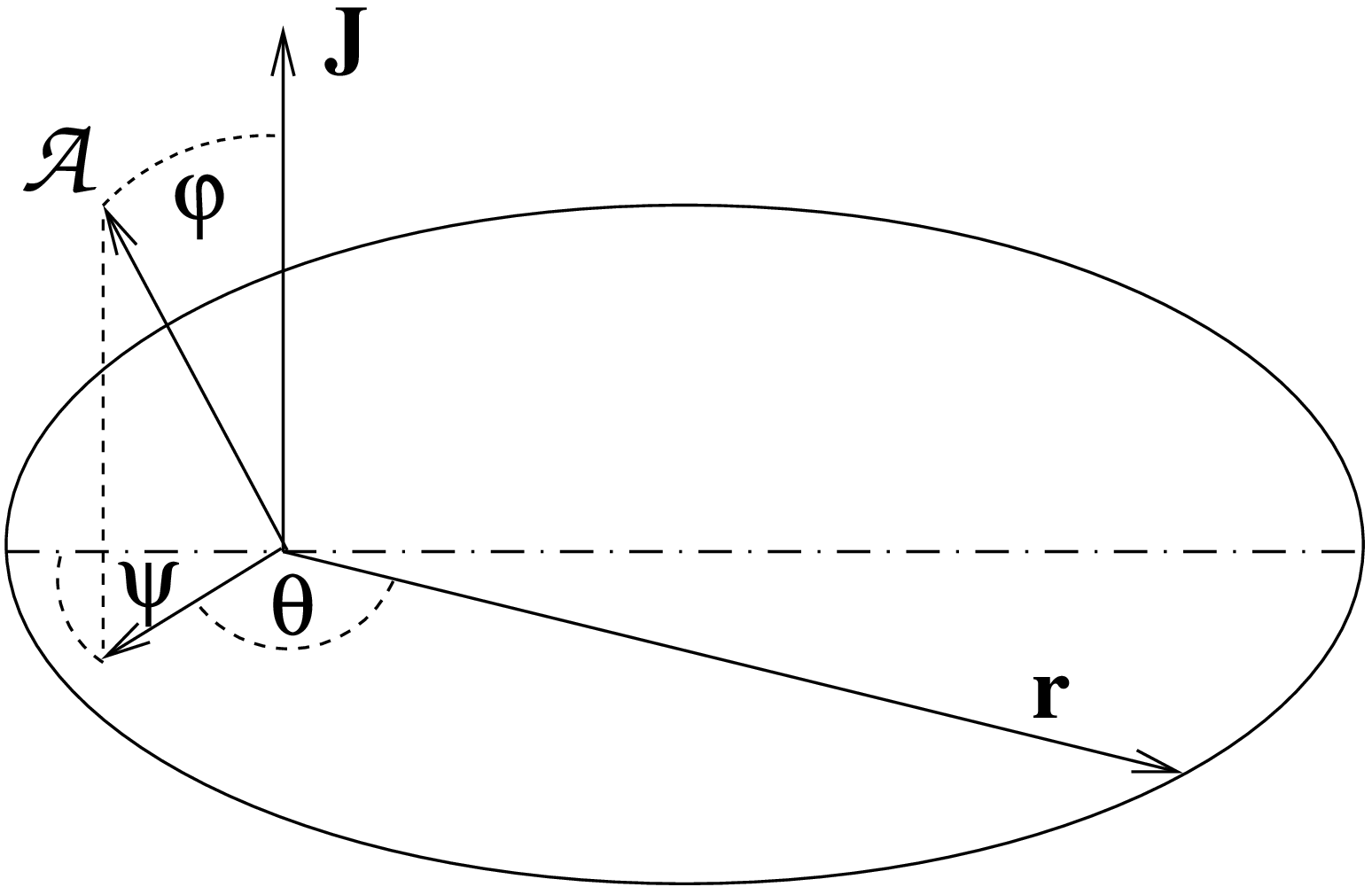}}
\end{center}
\figcaption{\label{fig:kepler}
Keplerian orbit of a star around the MBH binary.  The binary is centered 
at the root of the angular momentum vector in the drawing.}
\end{inlinefigure}

The effect of the Brownian motion
in galaxies could be enhanced if clumpy inhomogeneities are present in the
galactic potential. \citet{Backer:99} point out that the mass of
inhomogeneities at the center of the Milky Way varies with the projected
distance from position of Sgr A$^*$, assuming the distance of 8.5 kpc to the
Galactic center, as:
\beq
m(r)\sim 10^3 M_\odot \left(\frac{r}{\textrm{pc}}\right)^{1.5} .
\eeq
The accelerations implied by these inhomogeneities, 
$\accel\sim Gm(r)/r^2\approx
10^{-6}\textrm{ cm s}^{-1}$ at $r=1$ pc,
are comparable to the accelerations due to the stars that
are given by equation (\ref{eq:accelbrown}) and amount to
$\accel_*=0.5-3\times10^{-6}\textrm{ cm s}^{-1}$ using $m_*=M_\odot$,
$M_\bullet=2.6\times10^6M_\odot$, and $\sigma =75-110\textrm{ km s}^{-1}$.  
The presence of {\it compact}
inhomogeneities that do not suffer tidal truncation in the galactic potential,
such as intermediate-mass black holes, could in principle enhance the
amplitude of the Brownian motion beyond the predictions of this section.

Scaling the results of this section to the simulations of
\citet{Milosavljevic:01}, we find that the time scale 
on which the Brownian motion
fed stars into the loss cone amounted to 
$t_{\rm brown}\sim 1\textrm{ Myr}\times K(x)$,
which is of the order of the dynamical time.
While we argued above (\S~\ref{sec:nbody}) 
that gravitational scattering
into the loss cone was sufficient to keep it full in the $N$-body 
simulations, Brownian motion probably also contributed substantially
to the maintenance of a full loss cone in these simulations.
This underscores a critical difference between the published
$N$-body simulations and real galaxies; the $N$-body simulations are
dominated by small-$N$ effects and real galaxies are not.

\section{Coalescence}
\label{sec:coalescence}
When the MBHs approach each other sufficiently closely,
emission of gravity
waves becomes the dominant mechanism extracting the binding energy.
The binary
coalesces within a time $t_{\rm gr}$ when its semimajor axis $a_{\rm
gr}$
and the
eccentricity $e_{\rm gr}$ satisfy the relation derived by
\citet{Peters:64}:
\beq
a_{\rm gr}^4=\frac{64}{5}\frac{G^3M_1M_2(M_1+M_2) F(e_{\rm gr})}{c^5}
t_{\rm gr},
\eeq
where $F(e)$ has the form:
\beq
F(e)=\left(1-e^2\right)^{-7/2}
\left(1+\frac{73}{24}e^2+\frac{37}{96}e^4\right) .
\eeq
The ratio between the hard binary separation $a_{\rm
hard}=G\mu/4\sigma^2$
at which MBH binaries first form, and the
coalescence separation $a_{\rm gr}$ for circular binaries reads:
\bea
\label{eq:factorgr}
\frac{a_{\rm hard}}{a_{\rm gr}}
&\approx& 34 \times e^{9.2/\alpha}
\left(\frac{M_\bullet}{10^6M_\odot}\right)^{1/4-2/\alpha}\nonumber\\
& &\times
p^{3/4}(1+p)^{-3/2}\left(\frac{t_{\rm gr}}{10^9\textrm{ yr}}\right)^{-1/4} 
\eea
where, again, $\alpha$ is the slope of the $M_\bullet$--$\sigma$
relation, and
$p\equiv M_2/M_1\leq1$ is the ratio of the MBH masses.
The factor of $\sim 10-100$ that the MBH binary
needs to decay on its path from initial formation to coalescence is
sometimes called the ``final parsec problem.''
The required decay factor decreases
with decreasing mass ratio;
however when $p\lesssim 10^{-3}$, the smaller
MBH's host galaxy is likely to be severely
tidally disrupted during
the inspiral and thus deposit the black hole on an orbit
larger than $a_{\rm hard}$.
The time scale on which the smaller MBH spirals
inward due to dynamical friction could then be too
long to permit the formation of a hard binary
\citep{mer00,Yu:02a}
and the MBHs would not coalesce
even if the mechanisms discussed in this paper imply that
they should.

We proceed to a qualitative discussion of the scenarios for the evolution of 
MBH binaries with masses lying within several ranges of values.  
Bright, early type galaxies ($M_V < -21$) typically have shallow
nuclear density profiles,
$|d\log\nu/d\log r| \lesssim 1$, within a break radius $r_b$
 (e.g.,~\citealt{geb96}), 
although some exceptions to this rule have been found (e.g.,~\citealt{Laine:03}).  
The break radius is usually located well outside the dynamical radius of influence of the MBH.  
Therefore MBH binaries with initial black hole masses 
$M_\bullet\gtrsim 10^8M_\odot$ are from the outset situated in 
low-density stellar cusps and the mass initially inside the loss cone
is small compared with $M_\bullet$.
In addition, the time scales for diffusion into the loss cone are very 
long ($t_{\rm diff}\gtrsim 10^{11}$ yr).  
Therefore, binaries that form in the mergers of galaxies with black hole masses above $10^8M_\odot$ have difficulty decaying much beyond
the orbital separation of a hard binary.  
In this mass range, 
$a_{\rm hard}/a_{\rm gr}\sim 10$, implying that 
these binaries would be long-lived.
Exceptions might be merger events with extreme mass ratios,
$p\approx 10^{-2}$, for which the required decay factors are smaller.
If collisional loss cone refilling is very inefficient,
the dominant loss cone replenishment mechanism might be via
chaotic or centrophilic orbits if the potential is 
appreciably non-axisymmetric
\citep{Merritt:03}.  Even in axisymmetric, non-spherical potentials, 
larger stellar masses are available inside the loss cone \citep{Yu:02a} 
than suggested by the numerical estimates of the preceding sections 
(generalization of the results 
\S \ref{sec:timedep}--\S \ref{sec:brownian} to axisymmetric potentials
would be straightforward but tedious).
Furthermore, the longer a binary persists in an uncoalesced state,
the more likely that a third MBH will fall in during a subsequent merger,
facilitating decay by three-body interactions 
\citep{Makino:94,Valtonen:96}.
Thus we would still expect coalescence to sometimes occur
for high-mass binaries.

At the other end of the mass range, 
$M_\bullet\lesssim\textrm{ few}\times 10^6 M_\odot$, the black holes are expected to be embedded in approximately isothermal stellar cusps, 
$|d\log\nu/d\log r|\sim 2$.  
The formidable factor that the binaries need to cover from formation to 
coalescence, $a_{\rm hard}/a_{\rm gr}\sim 100$ is slightly
larger than for more massive binaries due to the
$M_\bullet$-dependence in equation (\ref{eq:factorgr}).
But a combination of processes contribute to effective orbital decay
in these galaxies.  
Foremost, $N$-body simulations \citep{Milosavljevic:01} have shown that the binaries shrink by a factor of $\sim 5-10$ beyond the hard binary separation immediately following the formation of a binary.  This takes place within a dynamical time and is a consequence of the abundance of cold (large binding energy) stars in the nascent loss cone to which the kinetic energy of the MBHs can be effectively transferred in slingshot ejections (see \S \ref{sec:newj}).  The ejection of most bound stars leads to a reduction of the density profile close to the sphere of influence of the MBH.  However there are good reasons to believe that galaxies with $M_\bullet\sim 10^6 M_\odot$ are in position to repair cusp damage on a time scale comparable to the binary hardening time scale, since both time scales are proportional to the relaxation time.  Indeed, the phase space transport due to two-body relaxation will tend to rebuild a cusp profile $\rho\propto r^{-7/4}$ around the MBH \citep{baw76}.  Moreover, the nearest and the best resolved MBH nucleus---Sgr A$^*$ cluster at the center of the Milky Way---exhibits evidence for young, massive stars of spectral type O-B deep inside the radius of influence of the MBH (e.g.,~\citealt{Figer:00,Ghez:03}).  This suggests that even in galaxy nuclei with $M_{{\rm Sgr A}^{*}}\sim 4\times10^6M_\odot$ black holes, star formation can repopulate the central cusp in as short a time as $10^7$ yr.

After the loss cone is empty of its original content, two-body relaxation lets stars diffuse into the loss region.  Note that the binary is still a factor $\sim 10-20$ wider than the separation conducive to gravitational coalescence in a Hubble time.  
From equation (\ref{eq:fluxtimeind}),
the equilibrium diffusion supplies an influx of stars at the rate:
\beq
\label{eq:mlostdiff}
\frac{dM_{\rm lost}}{dt}\sim \frac{3 M_\bullet}{10^{10}\textrm{ yr}} 
\left(\frac{M_\bullet}{10^6 M_\odot}\right)^{-1} \left(\frac{m_*}{M_\odot}\right) ,
\eeq
and the non-equilibrium diffusion could in principle supply stars at a rate larger by a factor of $\flux_{\rm noneq}/\flux_{\rm eq}\sim 2-10$. Combining equations (\ref{eq:mlostdiff}) and (\ref{eq:reejectlaw}), setting $m_*=M_\odot$, and using $\Delta\Phi/2\sigma^2\sim 10$ which is reasonable for isothermal galaxies with $M_\bullet\sim 10^{-3} M_{\rm gal}$, we find:
\bea
\label{eq:decaydiff}
\frac{d\ln a^{-1}}{dt}&\sim& 3\times 10^{-10}\textrm{ yr}^{-1} 
\left(\frac{a}{0.1 a_{\rm hard}}\right) \nonumber\\& &\times
\left(\frac{M_\bullet}{10^6 M_\odot}\right)^{-1} 
\left(\frac{\flux_{\rm noneq}}{\flux_{\rm eq}}\right) ,
\eea
where we have explictly indicated that diffusion becomes the dominant driver of binary hardening only after the immediate post-merger shrinkage 
to $a\sim 0.1 a_{\rm hard}$ has been completed.  The factor describing the enhancement of the loss cone flux out of equilibrium, $\flux_{\rm noneq}/\flux_{\rm eq}$, is greatly uncertain and large ($\sim 10$) when the phase-space is subject to episodic perturbations excited by the substructure inside the galaxy (orbiting giant molecular clouds, super-star clusters, etc.) which can 
restore the large phase-space gradients at the loss cone boundary (see \S~\ref{sec:timedep}).  We integrate equation (\ref{eq:decaydiff}) to obtain:
\beq
\frac{0.1 a_{\rm hard}}{a} \sim \frac{t}{3\times 10^9\textrm{ yr}} \left(\frac{M_\bullet}{10^6 M_\odot}\right)^{-1}\left(\frac{\flux_{\rm noneq}}{\flux_{\rm eq}}\right) .
\eeq
Although this result suffers from a variety of uncertainties, 
it suggests that for $M_\bullet\lesssim\textrm{ few}\times10^6M_\odot$, the binary separation can decay by a factor of at least a few out of the factor of
$\times 10-20$ that remain toward gravitational coalescence.
Hence we expect that stellar-dynamical processes might often serve
to bring MBH binaries in this mass range to complete coalescence.

Finally, MBHs in the intermediate range of 
$10^{6.5} M_\odot\lesssim M_\bullet\lesssim 10^8M_\odot$
live in nuclei with a range of cusp slopes, $0.5<|d\log\nu/d\log r| < 2$
\citep{geb96}.
The mechanisms discussed here would facilitate coalescence in galaxies
with steeper cusps, while in galaxies with shallower cusps, binaries
might be shrunk to separations within a factor of a few of $a_{\rm gr}$,
where other mechanisms (triaxiality, gas dissipation etc.) 
could plausibly bridge the remaining gap.

\section{Summary and Discussion}
We conclude with a tabulated summary of physical regimes that have
been discussed so far (Table \ref{tab:regimes}).

\begin{deluxetable}{lll}
\tablecolumns{3}
\tablewidth{30pc}
\tablehead{
  \colhead{} &
  \colhead{Form of Decay} &
  \colhead{Regime}}
\tablecaption{Physical Regimes for Long-Term Decay of 
Massive Black Hole Binaries}
\startdata
$a^{-1} \propto$ & $t + \textrm{const}$ & pinhole \\
$ $ & $t/N^\alpha + \textrm{const}$,\ \ \ \  
($0<\alpha<1$) & pinhole \& diffusion \\
$ $ & $t/N$ & diffusion \\
$ $ & $\textrm{const}$ & large-$N$ limit \\
$ $ & $\ln(1 + t/t_0) + \textrm{const}$ & pure reejection \\
$ $ & $\sqrt{1+Dt}$ & pinhole with reejection 
\enddata
\label{tab:regimes}
\end{deluxetable}

Clearly the long-term evolution of a massive black hole binary
can be very different in different environments.
We distinguish three characteristic regimes.

1. {\it Collisional.} The relaxation time is shorter than the
lifetime of the system and the phase-space gradients at the edge of
the loss cone are given by steady-state solutions to the Fokker-Planck
equation (\S~\ref{sec:reviewck}).
The densest galactic nuclei may be in this regime.
Resupply of the loss cone takes place on the time scale
associated with scattering of stars onto eccentric orbits.
Most of this scattering occurs in the ``diffusive''
(as opposed to ``pinhole'') regime and the decay
time of a MBH binary scales as $|a/\dot a|\sim m_*^{-1}\sim N$.
In the densest galactic nuclei, collisional loss cone
refilling may just be able to drive a MBH binary to coalescence
in a Hubble time (e.g.~\citealt{Yu:02a}).
$N$-body studies with any feasible $N$ are guaranteed to be
in the collisional regime, and may also be in the pinhole
regime corresponding to an always-full loss cone,
leading to artifically high decay rates, $a^{-1}\sim t$ (\S~\ref{sec:nbody}).
In $N$-body studies with $N\lesssim 10^6$, Brownian motion of the
binary further enhances the decay (\S~\ref{sec:brownian}).

2. {\it Collisionless.} The relaxation time is longer than
the system lifetime and gravitational encounters between
stars can be ignored.
Examples are the low-density nuclei of bright elliptical galaxies;
dark matter density cusps are also in this regime if the dark
matter particles are much less massive than stars.
The MBH binary quickly interacts with stars whose pericenters
lie within its sphere of influence; in a low-density nucleus,
the associated mass is less than that of the MBH binary and 
the decay tends to stall at a separation too large for gravity wave
emission to be effective.
However evolution can continue due to reejection of stars
that lie within the binary's loss cone but have not yet escaped from
the system.
In the spherical geometry, reejection implies $|a/\dot a|\sim 
(1+t/t_0)/a$, leading to a logarithmic dependence of binary
hardness on time (\S~\ref{sec:secondary}).
Reejection in galactic nuclei may contribute a factor of
$\sim$ a few to the change in $a$ in MBH binaries over a Hubble time.

3. {\it Intermediate.} The relaxation time is of order the age of the system.
While gravitational encounters contribute to the re-population of the
loss cone, not enough time has elapsed for the phase space
distribution to have reached a collisional steady state.
We argued that most galactic nuclei are likely to be in this regime.
The flux of stars into the loss cone can be substantially higher
than predicted by the steady-state theory, due to strong
gradients in the phase space density near the loss cone boundary
produced when the MBH initially formed (\S~\ref{sec:timedep}).
This transitory enhancement would be most important in a nucleus that continues
to experience mergers or infall, in such a way that the loss
cone repeatedly returns to an unrelaxed state with its associated 
steep gradients.

The evolution of a real MBH binary may reflect a combination of the
different regimes summarized above,
as well as other mechanisms that we have discussed only briefly 
or not at all.
We note a close parallel between the ``final parsec problem''
and the problem of quasar fueling: both requre that of order 
$10^8M_\odot$ be supplied to the inner parsec of a galaxy
in a time shorter than the age of the universe.
Nature clearly accomplishes this in the case of quasars,
probably through gas flows driven by torques from stellar bars
\citep{Begelman:90}. 
The same inflow of gas could contribute to the decay of a MBH
binary in a number of ways: by leading to the renewed formation
of stars which subsequently interact with the binary;
by inducing torques which extract 
angular momentum from the binary \citep{Armitage:02};
through accretion, increasing the masses of one or both
of the MBHs and reducing their separation; etc.

In massive elliptical galaxies or spheroids, the fraction
of mass in the form of gas during the last major merger
event is likely to have been small \citep{kah00}.
In addition to the mechanisms discussed in this paper,
decay of a MBH binary could be enhanced by the presence
of a significant population of ``centrophilic'' orbits,
orbits that pass near the center of the potential once per
crossing time.
Such orbits may constitute a large fraction of the mass of
a triaxial nucleus \citep{Poon:02}.
Even in galaxies where none of these mechanisms is effective and
the decay stalls, infall of a third
MBH following a merger or accretion event could accelerate
the decay via the Kozai mechanism \citep{Blaes:02}, 
or the strong three-body gravitational interactions 
\citep{Makino:94,Valtonen:96},
which induce large changes in the binary's eccentricity
and enhanced rates of gravity wave emission at pericenter.

We have not attempted in this paper to make quantitative
estimates of the decay rates or stalling radii of MBH binaries
in real galaxies.
Such a program would be problematic for a number of reasons,
most importantly the uncertain history of the loss cone
and the unknown elapsed time and character of the most recent merger
(\S~\ref{sec:timedep}).
However we argued that a recent study \citep{Yu:02a}
could have overestimated stalling radii,
due both to the neglect of the various new mechanisms discussed here,
and also by ignoring the influence that MBH formation must
have had on the form of the nuclear density profile:
this  profile must have been steeper before the MBH binary formed,
leading to substantially enhanced loss cone fluxes and more
rapid decay early in the life of the binary.
Nevertheless, none of our results would lead us to rule out
the existence of uncoalesced MBH binaries in at least some galaxies,
particularly galaxies with large, low-density nuclei and
a little gas (\S~\ref{sec:coalescence}).

A striking and disappointing conclusion of this study is the
difficulty of using numerical $N$-body simulations to 
follow the long-term evolution of MBH binaries (\S~\ref{sec:nbody}).
We argued that a variety of collisional effects associated with
small $N$ would lead to a more rapid decay of the binary than
in real galaxies.
While there is still a useful role for $N$-body simulations
in following the initial formation and early evolution of a MBH binary,
when much of the decay and associated cusp destruction takes place,
the rapid decay of $a(t)$ seen in most published simulations is 
due to loss-cone repopulation occuring at rates much higher
than expected in real galaxies.
We believe that the future role of $N$-body simulations
in this field will be limited to predicting the form of the
phase-space gradients produced during the formation of a MBH binary,
which can then be adapted as initial conditions for a collisionless
or Fokker-Planck treatment.
A carefully designed set of $N$-body experiments might also
be used to understand the $N$-dependence of collisional loss
cone effects which could then be extrapolated to the larger-$N$
regime characteristic of real galaxies.

\vspace{0.5cm}
We are grateful to
A. Kosowsky, C. Pryor, and R. Spurzem for their detailed
comments on an early version of the manuscript.  We acknowledge helpful
discussions with J. Binney, S. Phinney, F. Rasio, and M. Rees.
This work was supported by NSF grants AST 96-17088 and AST 00-71099
and by NASA grants NAG5-6037 and NAG5-9046 and STScI grant HST-AR-08759.
M.~M.\ thanks the
Sherman Fairchild Foundation for support.

\end{document}